\PassOptionsToPackage{table}{xcolor}
\documentclass[sigconf,screen]{acmart}

\newif\ifpreprint
\preprinttrue   

\usepackage{tabularx}
\usepackage{dcolumn} 
\newcolumntype{d}[1]{D{.}{.}{#1}}
\usepackage{acmart-taps}
\usepackage{subcaption}
\usepackage{enumitem}
\usepackage{colortbl}
\usepackage{array} 
\usepackage{framed}

\usepackage{makecell}

\usepackage{bm}
\usepackage{booktabs} 
\usepackage{graphicx}
\usepackage{caption}
\usepackage{rotating}
\usepackage{listings}
\usepackage{upquote}   
\usepackage{multirow}
\usepackage{placeins}

\lstdefinestyle{python}{
  language=Python,
  basicstyle=\ttfamily\footnotesize, 
  numbers=left,
  numberstyle=\tiny,
  stepnumber=1,
  numbersep=6pt,
  showstringspaces=false,
  columns=fullflexible,
  keepspaces=true,
  upquote=true,
  frame=single,
  framerule=0.3pt,
  breaklines=true,
  breakatwhitespace=true,
  postbreak=\mbox{\textcolor{gray}{$\hookrightarrow$}\space},
  tabsize=2,
  aboveskip=4pt,
  belowskip=4pt,
}
\definecolor{textcond}{HTML}{55A868} 
\definecolor{visualcond}{HTML}{E2825B} 
\definecolor{combicond}{HTML}{8172B2} 
\definecolor{textcond20}{RGB}{221,238,225} 
\definecolor{visualcond20}{RGB}{249,230,222} 
\definecolor{combicond20}{RGB}{230,227,240} 

\definecolor{textcolor}{HTML}{72B6A1}
\definecolor{visualcolor}{HTML}{E99675}
\definecolor{combicolor}{HTML}{95A3C3}
\definecolor{hellblue}{HTML}{8EC8FF}

\newif\iftaps
\tapsfalse   

\iftaps
    \renewcommand{\textonly}[1]{#1}
    \renewcommand{\visualonly}[1]{#1}
    \renewcommand{\combionly}[1]{#1}
    \renewcommand{\highlight}[1]{#1}
\else
    \newcommand{\highlight}[1]{\cellcolor{gray!40}{#1}}
    \newcommand{\textonly}[1]{\cellcolor{textcond20}{#1}}
    \newcommand{\visualonly}[1]{\cellcolor{visualcond20}{#1}}
    \newcommand{\combionly}[1]{\cellcolor{combicond20}{#1}}
\fi

\newcolumntype{P}{>{\centering\arraybackslash}m{1.6em}} 

\AtBeginDocument{%
  }

\copyrightyear{2026}
\acmYear{2026}
\acmConference[IUI '26]{31st International Conference on Intelligent User Interfaces}{March 23--26, 2026}{Paphos, Cyprus}
\acmBooktitle{31st International Conference on Intelligent User Interfaces (IUI '26), March 23--26, 2026, Paphos, Cyprus}

\ifpreprint
    \setcopyright{none}
    
    \renewcommand\footnotetextcopyrightpermission[1]{%
        \footnotetext{%
        \small\textit{Preprint. Accepted for publication in IUI '26. The final published version will appear in the ACM Digital Library.}%
        }%
    }

\else
    \setcopyright{cc}
    \setcctype{by}

\fi

\begin{document}

\title[Enhancing Generative AI Image Refinement with Scribbles and Annotations]{Enhancing Generative AI Image Refinement with Scribbles and Annotations: A Comparative Study of Multimodal Prompts}

\author{Hyerim Park}
\orcid{0009-0006-4877-2255}
\affiliation{%
  \institution{BMW Group}
  \city{Munich}
  \country{Germany}}
\affiliation{
  \institution{University of Stuttgart}
  \city{Stuttgart}
  \country{Germany}}
\email{hyerim.park@bmw.de}

\author{Phuong Thao Tran}
\orcid{0009-0006-9390-1714}
\affiliation{%
  \institution{BMW Group}
  \city{Munich}
  \country{Germany}}
\affiliation{%
  \institution{TU Berlin}
  \city{Berlin}
  \country{Germany}}
\email{tran.2@campus.tu-berlin.de}

\author{Andre Luckow}
\orcid{0000-0002-1225-4062}
\affiliation{%
  \institution{BMW Group}
  \city{Munich}
  \country{Germany}}
\affiliation{
  \institution{LMU Munich}
  \city{Munich}
  \country{Germany}}
\email{andre.luckow@bmwgroup.com}

\author{Ceenu George}
\orcid{0009-0002-8616-0234}
\affiliation{%
 \institution{TU Berlin}
 \city{Berlin}
 \country{Germany}}
\email{ceenu.george@tu-berlin.de}

\author{Michael Sedlmair}
\orcid{0000-0001-7048-9292}
\affiliation{%
 \institution{University of Stuttgart}
 \city{Stuttgart}
 \country{Germany}}
\email{michael.sedlmair@visus.uni-stuttgart.de}

\author{Malin Eiband}
\orcid{0000-0003-4024-1645}
\affiliation{%
  \institution{BMW Group}
  \city{Munich}
  \country{Germany}}
\email{malin.eiband@bmw.de}

\renewcommand{\shortauthors}{Park et al.}

\begin{abstract}
Generative AI (GenAI) image tools are increasingly used in design practice, enabling rapid ideation but offering limited support for refinement tasks such as adjusting layout, scale, or visual attributes. While text prompts and inpainting allow localized edits, they often remain inefficient or ambiguous for precise, in-context, and iterative refinement---motivating the exploration of alternative methods. This work examines how pen-based scribbles and annotations can enhance GenAI image refinement. A formative study with seven professional designers informed a prototype supporting three input modalities: text-only, visual-only, and combined prompting. A within-subjects study with 30 designers and design students compared these modalities across closed- and open-ended tasks, evaluating expressiveness, efficiency, workload, user experience, iteration, and multimodal strategies. Visual prompts improved clarity and speed for spatial edits while reducing workload, whereas text remained effective for semantic and global changes. The combined modality received the highest overall ratings, enabling complementary use, balancing spatial precision with semantic detail, and supporting smoother iteration. Task-specific preferences also emerged: adding new objects often required both modalities, while moving or modifying elements was typically handled through visual input. This work contributes (1) an empirical comparison of multimodal prompting for GenAI refinement, (2) a prototype integrating scribbles and annotations, and (3) insights into designers' multimodal strategies to inform future GenAI interfaces that better support refinement in GenAI-supported design workflows.
\end{abstract}



\begin{teaserfigure}
    \centering
    \includegraphics[width=\linewidth]{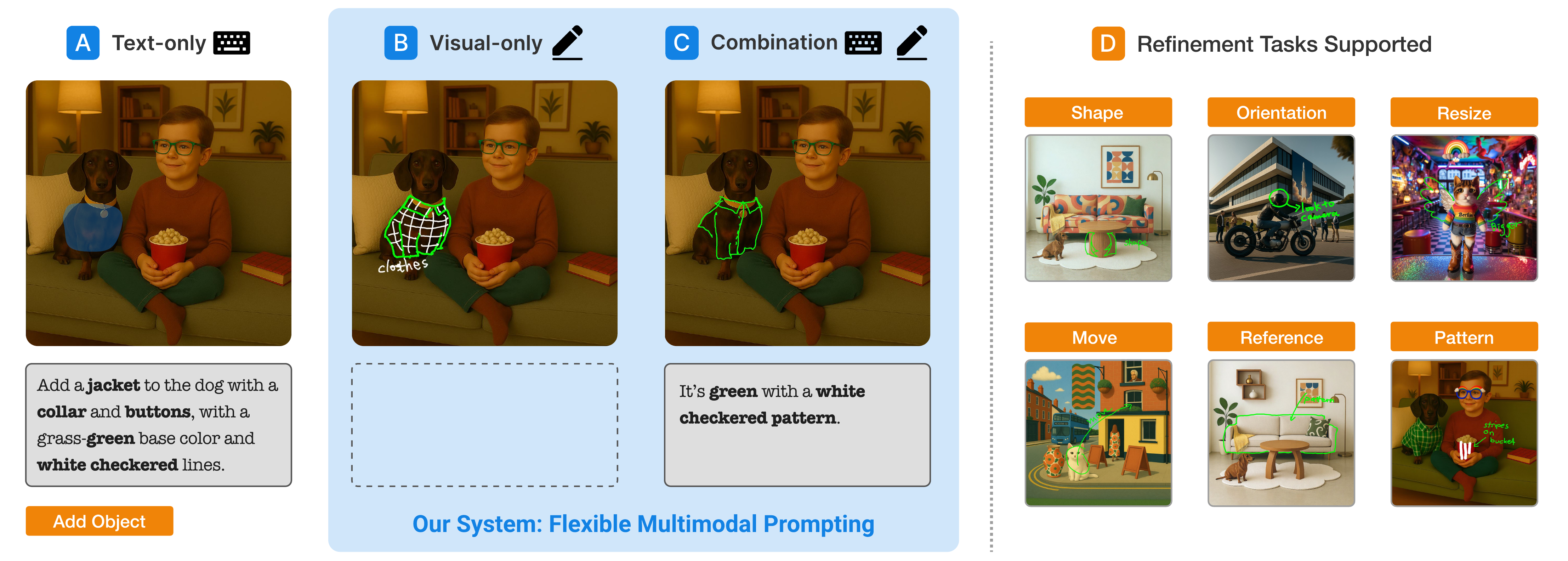}
    \caption{\emph{Multimodal prompting for GenAI image refinement.} The system supports (A) text-based input with inpainting, (B) pen-based scribbles and annotations as visual input, and (C) combined multimodal input, enabling users to perform (D) various refinement tasks---from spatial edits (e.g., move, resize, reorient) to semantic or stylistic changes (e.g., add or modify patterns)---that balance precision and expressiveness.}
    \label{fig:teaser}
    \Description{The teaser figure illustrates a system for multimodal prompting in GenAI image refinement. Panel A shows text-only input, where a user specifies an edit using natural language to modify an image region. Panel B shows visual-only input using pen-based scribbles and annotations drawn directly on the image. Panel C shows combined input, where text and visual annotations are used together to refine a specific object. Panel D presents examples of supported refinement tasks, including changes to shape, orientation, size, position, reference objects, and visual patterns. Together, the panels demonstrate how the system supports flexible image refinement through text, visual input, and their combination.}
\end{teaserfigure}

\maketitle

\section{Introduction}
Generative AI (GenAI) image tools\footnote{In this paper, ``GenAI'' refers specifically to generative AI for image generation.} such as Midjourney~\cite{midjourney_midjourney_2025}, DALL-E~\cite{dalle_3_dalle_2025}, and Adobe Firefly~\cite{adobe_adobe_2025} enable the creation of high-quality images from various inputs, such as text, images, and sketches, and often produce results comparable to human-produced visuals. These systems have been rapidly adopted across diverse design domains, including product, industrial, graphic, UI/UX, and automotive design~\cite{liu_3dall-e_2023, feng_how_2023, chen_autospark_2024, son_genquery_2024, liu_opal_2022, brade_promptify_2023, kim_shoegenai_2025, park_design_2025}. They are particularly effective for rapid ideation, allowing designers to quickly generate variations and explore creative directions~\cite{wang_aideation_2025, choi_creativeconnect_2024, sun_creative_2025, wang_roomdreaming_2024, verheijden_collaborative_2023}. However, current GenAI tools remain less suited for refinement tasks, where designers iteratively adjust image outputs to match their intent. Such refinements often involve localized, in-context edits, such as adjusting spatial composition, scale, orientation, referencing objects within the image, or modifying specific visual attributes while preserving other elements. Expressing these nuanced intentions through text alone with inpainting can be inefficient and ambiguous, creating a gap between designers' predominantly visual ways of working and the text-based interfaces of most GenAI systems~\cite{park_we_2024, gmeiner_exploring_2023, lin_inkspire_2025}. Without adequate support for refinement, GenAI image tools may remain limited to early-stage ideation rather than becoming integrated into other stages of design workflows~\cite{shi_brickify_2025, dang_worldsmith_2023, lin_sketchflex_2025, wang_gentune_2025}. 

Prior work has mainly explored two directions to address these limitations. The first focuses on improving text prompting by introducing approaches such as keyword suggestions, automatic rephrasing, and visualizations of text--output relationships to reduce trial--and--error~\cite{brade_promptify_2023, wang_promptcharm_2024, adamkiewicz_promptmap_2025}. The second investigates multimodal interaction, combining text with sketches, segmentation masks, or color cues to convey intent more directly and visually~\cite{lin_inkspire_2025, lin_sketchflex_2025, shi_brickify_2025}. These studies show that visual input can bridge the gap between designers' intent and system interpretation, allowing users to express ideas more intuitively and improve communication with GenAI systems during image generation and refinement. However, while scribbles and annotations are already common in design practice as quick, expressive tools for externalizing and sharing ideas, their potential as actionable prompts for refinement---rather than ideation---remains underexplored.

This work investigates how pen-based scribbles and annotations can enhance GenAI image refinement. We define scribbles not as mere sketches but as actionable visual prompts that communicate refinement intent, such as moving an object, resizing it, or transferring a pattern. To explore this, we first conducted a formative study with seven professional designers to understand how text, scribbles, and annotations support the expression of refinement in GenAI. Insights from this study informed the design of a prototype that supports three input modalities: (1) text-only prompting with inpainting, (2) visual prompting through pen-based scribbles and annotations, and (3) a combined modality integrating both. We then conducted a within-subjects study with 30 designers and design students across closed- and open-ended refinement tasks, aiming to address the following research questions:
\aptLtoX{\begin{enumerate}
    \item[RQ1] How do text-only, visual-only, and combined prompting differ in supporting the \textbf{expression of refinement intent}?
    \item[RQ2] How do the three input modalities compare in \textbf{efficiency}, \textbf{workload}, and \textbf{user experience} during refinement?
    \item[RQ3] How do users \textbf{iteratively} refine and evolve images across the three input modalities?
    \item[RQ4] When both inputs are available, how do users \textbf{combine or switch} between input modalities, and what \textbf{strategies} emerge?
\end{enumerate}}{\begin{enumerate}[label=\textbf{RQ\arabic*}, leftmargin=*]
    \item How do text-only, visual-only, and combined prompting differ in supporting the \textbf{expression of refinement intent}?
    \item How do the three input modalities compare in \textbf{efficiency}, \textbf{workload}, and \textbf{user experience} during refinement?
    \item How do users \textbf{iteratively} refine and evolve images across the three input modalities?
    \item When both inputs are available, how do users \textbf{combine or switch} between input modalities, and what \textbf{strategies} emerge?
\end{enumerate}}

Our findings show that visual prompts improved clarity for spatial and relational edits, reduced workload, and enabled faster completion of spatial tasks compared to text-only input. In contrast, text prompts remained effective for semantic and global refinements, such as specifying object type, material, or overall style, where language provided precision by leveraging the model's conceptual and domain knowledge. The combined modality received the highest overall ratings, supporting both complementary strategies (e.g., sketching spatial intent and specifying details in text) and redundant strategies (repeating information across modalities to reduce ambiguity). Participants frequently adopted ``visual-first'' workflows and layered modalities to increase confidence and control during iteration. Task-specific patterns also emerged: adding new objects often required both modalities, whereas modifying patterns or moving objects was typically expressed with visual prompts alone.

Overall, these results indicate that visual prompting enables more direct and unambiguous communication of refinement intent, while text input remains valuable for describing semantic and stylistic properties that benefit from linguistic precision. By affording more precise, expressive, and intuitive refinements, multimodal prompting may better align GenAI tools with professional design practice. This paper contributes (1) an empirical comparison of text-only, visual-only, and combined prompting modalities for GenAI image refinement, (2) a prototype system that integrates pen-based scribbles and annotations with text-based inpainting into multimodal refinement workflows, and (3) insights into how designers adopt, combine, and switch between modalities, with implications for the design of future multimodal GenAI interfaces that better support refinement and creative design workflows.

\section{Related Work}
Research in HCI and design has explored how GenAI tools are integrated into creative workflows, their limitations with text-based prompting, and the potential of multimodal interaction. We review three strands of work most relevant to our study: (1) interfaces for GenAI image tools, (2) support for refinement tasks, and (3) the role of scribbles and annotations in design practice.

\subsection{GenAI Image Tool Interfaces}
In recent years, HCI research has examined how GenAI image systems are integrated into creative workflows and how their interfaces influence design practice. Studies highlight both opportunities and challenges: while GenAI enables rapid exploration of design concepts, designers often face the cognitive burden of crafting effective prompts, mismatches between intent and output, and repeated trial--and--error corrections that distract from core design activities~\cite{park_we_2024, chen_autospark_2024, vimpari_adapt-or-type_2023, gmeiner_exploring_2023}. Current systems typically interpret prompts in isolation, with limited contextual awareness of design intent or prior iterations~\cite{tholander_design_2023, shi_understanding_2023}, and remain optimized for single-step generation rather than the iterative and reflective nature of design work~\cite{huang_plantography_2024, li_realtimegen_2025, park_we_2024}.  

To address these limitations, researchers have proposed two main interface directions~\cite{park_designing_2024}. The first enhances \textbf{text prompting}. Researchers have developed systems to make prompt creation less burdensome, including keyword suggestions to inspire alternatives and reduce fixation~\cite{choi_creativeconnect_2024, peng_designprompt_2024, wang_aideation_2025}, dropdowns for combinatorial exploration~\cite{wang_promptcharm_2024}, and automatic rephrasing or augmentation for more consistent outputs~\cite{brade_promptify_2023, son_genquery_2024}. Visualization approaches such as PromptMap~\cite{adamkiewicz_promptmap_2025} cluster images by semantic similarity to help users understand how different keywords influence outputs. Together, these tools aim to reduce trial--and--error in prompt creation by supporting guided and interpretable processes. 
The second explores \textbf{multimodal interaction}, incorporating sketches, segmentation masks, reference images, and color palettes~\cite{peng_designprompt_2024, dang_worldsmith_2023, lin_inkspire_2025}. Such methods allow users to externalize intent more directly---for example, by drawing rough layouts~\cite{lin_inkspire_2025}, specifying textures~\cite{peng_designprompt_2024}, or uploading references~\cite{choi_creativeconnect_2024}. GUI-based controls (e.g., sliders, anchors) make refinements more interactive and interpretable~\cite{chung_promptpaint_2023, zhang_protodreamer_2024, michelessa_varifai_2025}. Analogy-driven methods use visual tokens for spatial relations~\cite{shi_brickify_2025}, pen-based attribute transfer and reference~\cite{peng_fusain_2025}, and color blocking for intuitive composition~\cite{lin_sketchflex_2025}. These approaches show the potential of multimodal input to better align GenAI image tools with design workflows.  
Building on this second direction, our work systematically compares text, visual, and combined prompting, evaluating their effects on expressiveness, efficiency, workload, user experience, and iterative refinement, as well as the strategies designers adopt when combining inputs.
 
\subsection{Refining Images in GenAI}
Most prior work on GenAI image systems has focused on ideation and divergent exploration, using AI to rapidly generate concepts and stimulate creativity~\cite{tholander_design_2023, chen_autospark_2024, choi_creativeconnect_2024, brade_promptify_2023, son_genquery_2024, liang_storydiffusion_2025, choi_expandora_2025}. These affordances fit early design phases such as brainstorming and concept exploration~\cite{tang_exploring_2024, park_we_2024, inie_designing_2023, wang_aideation_2025, chen_how_2025}. In contrast, \textbf{refinement}---where designers iteratively adjust images to better align with an evolving vision---has received far less attention. 
We define refinement as the iterative process of fine-tuning images to match a designer's intent, emphasizing local and contextual adjustments rather than full regeneration. It also differs from finalization, which focuses on pixel-perfect production deliverables and execution precision. As design progresses from broad stylistic exploration toward detailed spatial composition, the need for iterative, localized adjustments increases~\cite{dang_worldsmith_2023, zhang_protodreamer_2024}.

Existing tools predominantly support partial modification via inpainting or outpainting~\cite{yu_generative_2018, dalle_inpainting_editing_2024}. While these methods allow users to edit specific regions, they depend heavily on textual descriptions, which can be ambiguous when specifying spatial relations or subtle refinements~\cite{dang_worldsmith_2023, liang_storydiffusion_2025}. End-to-end generation often yields partially satisfactory results that still require correction~\cite{lin_sketchflex_2025, uusitalo_clay_2024, verheijden_collaborative_2023}. Designers frequently report frustration when modifying one aspect of an image while preserving others, as current workflows require repeated prompts for engineering or manual adjustments in external tools~\cite{lin_sketchflex_2025, park_we_2024}. To address these limitations, recent research has begun to explore alternative approaches. Systems such as Brickify~\cite{shi_brickify_2025} introduce visual tokens to specify spatial relationships; GenTune~\cite{wang_gentune_2025} links image elements to prompt labels for traceable refinements; and FusAIn~\cite{peng_fusain_2025} proposes pen-based interactions for transferring attributes from a specific image and reusing generated content. These efforts highlight growing recognition of refinement as a critical stage in GenAI image generation workflows. In line with this direction, we systematically compare text, visual, and combined input modalities to examine how scribbles and annotations function as actionable prompts for refinement during image generation and how multimodal interaction supports more iterative, precise design refinement.

\subsection{Scribbles and Annotations in Design Practices}
Sketching, scribbling, and annotating have long been central to design practice~\cite{schon_designing_1992, goldschmidt_backtalk_2003}, serving as cognitive and communicative tools for externalizing ideas and reasoning about spatial relationships~\cite{goldschmidt_backtalk_2003, plimmer_computer-aided_2002}. Quick and lightweight visual expressions such as arrows, circles, or short notes help designers convey structure, proportion, and intent without lengthy verbal descriptions, reducing cognitive load and facilitating collaboration~\cite{vogel_conte_2011, hinckley_pen_2010, jonson_design_2005}. The interpretive flexibility of sketching supports both exploratory and precise thinking~\cite{gross_ambiguous_1996}, and pen-based input has been shown to feel natural and expressive across a wide range of design tasks~\cite{hinckley_pen_2010, landay_interactive_1995}. Many designers also report preferring to express ideas visually rather than verbally~\cite{park_we_2024}.  

HCI research has extended these practices into GenAI-supported workflows. Systems such as Inkspire~\cite{lin_inkspire_2025} integrate abstract sketches to outline object shapes, SketchFlex~\cite{lin_sketchflex_2025} employs color blocking to refine composition, Brickify~\cite{shi_brickify_2025} encodes spatial relationships through visual tokens, and FusAIn~\cite{peng_fusain_2025} enables pen-based transfer of attributes across objects. These systems illustrate how visual input can effectively convey spatial intent, layout, or attribute references during iterative GenAI image generation and refinement. 

However, commercial GenAI platforms still prioritize text input, with visual input largely limited to masking or reference uploads~\cite{zamfirescu-pereira_why_2023}. Extensions such as ControlNet expand conditioning on posture, edges, or depth maps~\cite{lllyasviel_lllyasvielcontrolnet_2025}, but remain specialized pipelines. Comparative research suggests that while text excels at global ideation, sketches and scribbles are particularly effective for refinement tasks~\cite{lee_impact_2024}. Designers often move fluidly between these modalities, using visual input for local or spatial adjustments and text for broader semantic directions~\cite{son_genquery_2024}. Nevertheless, systematic empirical comparisons of text, visual, and combined input for refinement remain scarce. Our work explicitly examines how scribbles and annotations---common in design practice---serve as actionable prompts and are used within multimodal refinement workflows.

\section{Formative Study}
\label{sec:formative}
To understand how designers express refinement intentions using different input modalities and to inform the design of our prototype, we conducted a formative study with professional designers from diverse fields. The study examined how text prompts, annotations, and scribbles are used in image refinement within GenAI workflows and how designers envision integrating these inputs into GenAI-supported design practice.

\subsection{Participants}
Seven professional designers (3 female, 4 male) participated, with backgrounds in UI/UX, 3D/XR, graphic, and automotive interior design. Their experience with GenAI image tools---such as Midjourney, DALL-E, and Adobe Firefly---varied: one participant used such tools daily ($n = 1$), while others reported usage frequencies of two to three times weekly ($n = 2$), weekly ($n = 2$), monthly ($n = 1$), and less than monthly ($n = 1$).

\subsection{Study Design and Procedure}
We used a \textbf{digital paper-based prototype} implemented on an iPad with a stylus, keyboard, and mouse, intentionally excluding model generation to focus on participants' input behavior and preferences while avoiding bias from output quality or latency. Each in-person session lasted approximately 70 minutes and included four phases: (1) introduction and tutorial, (2) pre-tasks, (3) main refinement tasks, and (4) a semi-structured interview. 
Participants interacted with three input modalities:  
\begin{itemize}[leftmargin=*]
    \item \textbf{Text prompts:} typed instructions entered via keyboard.
    \item \textbf{Annotations:} handwritten text or symbols (e.g., arrows, circles, short labels) added directly on the image using a stylus or mouse.
    \item \textbf{Scribbles:} free-form sketches or marks drawn directly on the image using a stylus or mouse.  
\end{itemize}

For text prompts, participants typically specified edit regions using an \textit{inpainting} selection brush operated via a mouse, following conventions in commercial tools such as DALL-E and Midjourney. For annotations and scribbles, participants could directly mark edit areas (e.g., circles, arrows) on the image using the same pen tool, without requiring additional menu navigation; the selection brush could still be used if desired. 
Each participant completed six refinement tasks: (T1) adding a new element; (T2) modifying an existing element; (T3) revising material or surface patterns; (T4) making structural or spatial adjustments; (T5) applying global modifications; and (T6) performing a self-directed refinement. Task order followed a Latin square design. Participants freely chose their preferred input modalities for each task and could select multiple modalities when needed. Think-aloud protocols, observation notes, and post-session interviews captured participants' perceived strengths and limitations and their reflections on modality use.

\subsection{Findings}
Qualitative analysis revealed distinct strengths and limitations for each modality, as well as designers' tendencies to combine them---for example, combining scribbles and annotations.

\textbf{Text Prompts.} 
Participants preferred text when they had a clear mental image of the desired outcome or needed to specify multiple attributes precisely in one prompt (e.g., \textit{``Add a Santa hat in wine-red velvet with sequin decorations''}). Text was also used when they wanted the AI to contribute creatively with minimal user-provided direction. It was deemed well-suited for global modifications, such as adjusting overall style or tone, but inefficient for spatial adjustments or localized changes. Several participants noted that describing visual ideas verbally was cognitively demanding.

\textbf{Annotations.} 
Annotations were regarded as particularly effective for spatial adjustments and referencing specific objects or attributes within the image. Designers used arrows, circles, and short notes to indicate movement, resizing, or attribute transfers between objects (e.g., applying a fabric texture from one object to another). Although quick and clear to use, excessive written annotations sometimes cluttered the image, and participants questioned whether AI systems could reliably interpret handwriting or shorthand.

\textbf{Scribbles.} 
Scribbles helped participants express object shape, size, and position directly on the image and were particularly valued for familiar or straightforward forms (e.g., ribbons, footballs). At the same time, participants appreciated their immediacy for quick ideation through rough sketches, but rarely relied solely on scribbles; they often combined them with annotations or text to clarify meaning.

\textbf{Combined Use.} 
Participants naturally switched between modalities or layered them during refinement. Combining scribbles and annotations enabled clear and direct visual communication, suggesting that these inputs could be merged into a single visual modality, as participants perceived little distinction between them. Pairing visual input with text further reduced ambiguity and balanced spatial precision with semantic detail. Designers described multimodal use as both efficient and creatively satisfying, viewing it as an intuitive and practical way to convey refinement intent. These observations directly informed the design of the interactive prototype and the comparative user study that followed.

\begin{figure*}[t] 
    \centering
    \includegraphics[width=\linewidth]{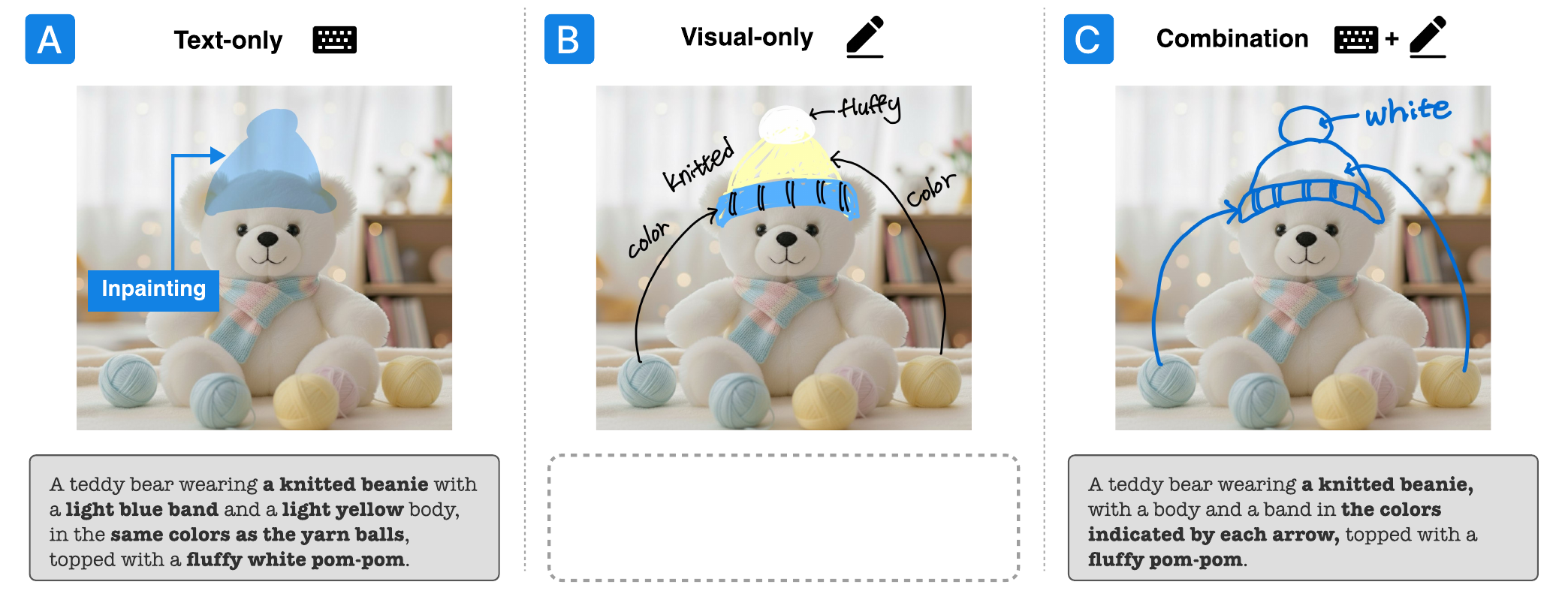} 
    \caption{Overview of the three input modalities in the prototype: 
    (A) \textbf{Text-only}, using typed prompts with inpainting-based region selection; 
    (B) \textbf{Visual-only}, using pen-based scribbles and annotations; and 
    (C) \textbf{Combination}, using multimodal input integrating both text and visual inputs, which are each available for use as needed.}
    \label{fig:input_methods}
    \Description{The figure presents three panels illustrating input modalities supported by the prototype. Panel A shows text-only input, where a user types a natural-language prompt and selects a target region through inpainting to modify part of an image. Panel B shows visual-only input, where a user draws pen-based scribbles and annotations directly on the image to indicate desired attributes such as shape or color. Panel C shows combined multimodal input, where text prompts and visual annotations are used together, but can also be applied independently, to specify refinements to the same object. The figure compares how text, visual, and combined inputs support image refinement.}
\end{figure*}

\subsection{Design Goals}
Insights from the formative study informed five design goals (DG1--DG5) for the subsequent prototype:
\aptLtoX{\begin{enumerate}
    \item[DG1] \textbf{Unified Visual Input:} Merge scribbles and annotations into one visual mode reflecting how designers naturally combined them.    
    \item[DG2] \textbf{Pen Interaction Support:} Provide intuitive pen-based input consistent with professional sketching practices.
    \item[DG3] \textbf{Enable Multimodality:} Allow flexible switching between and combination of text and visual inputs to accommodate diverse refinement needs and workflows.  
    \item[DG4] \textbf{Iterative Workflow Support:} Support incremental, history-based refinement rather than single-step generation.   
    \item[DG5] \textbf{Improve Interpretability:} Improve system recognition of visual marks, scribbles, and handwritten notes to minimize ambiguity and increase user confidence. 
\end{enumerate}}{\begin{enumerate}[label=\textbf{DG\arabic*}, leftmargin=*]
    \item \textbf{Unified Visual Input:} Merge scribbles and annotations into one visual mode reflecting how designers naturally combined them.    
    \item \textbf{Pen Interaction Support:} Provide intuitive pen-based input consistent with professional sketching practices.
    \item \textbf{Enable Multimodality:} Allow flexible switching between and combination of text and visual inputs to accommodate diverse refinement needs and workflows.  
    \item \textbf{Iterative Workflow Support:} Support incremental, history-based refinement rather than single-step generation.   
    \item \textbf{Improve Interpretability:} Improve system recognition of visual marks, scribbles, and handwritten notes to minimize ambiguity and increase user confidence. 
\end{enumerate}}

\section{Prototype Design and Implementation} 
\label{sec:prototype}
Building on the five design goals (DG1--DG5) derived from the formative study, we developed a web-based prototype supporting multimodal interaction for GenAI image refinement. The prototype enables designers to express refinement intent through text input, pen-based scribbles and annotations, or a combination of these, thereby supporting the operationalization of the design goals.

\subsection{Input Modalities}
The system provides three input modalities for image refinement (\autoref{fig:input_methods}): (1) \textbf{Text-only}, (2) \textbf{Visual-only}, and (3) \textbf{Combination}, a multimodal condition that supports flexible use of text and visual inputs.

\textbf{Text-only.} 
Users select regions to edit using a brush-based \textit{inpainting} tool operated with a mouse and enter a textual prompt (e.g., ``add a knitted beanie''). This workflow mirrors how text-based refinement is typically performed in current GenAI tools (e.g., DALL-E, Midjourney) and serves as a practice-grounded baseline condition for comparison. Observations from the formative study, together with conventions in existing commercial GenAI systems, informed this design choice: designers commonly associated text-based refinement with mouse-driven inpainting. Accordingly, the text-only condition reflects prevailing real-world GenAI workflows rather than enforcing uniform input device settings across conditions.

\textbf{Visual-only.}
Following \textbf{DG1} and \textbf{DG2}, scribbles and annotations are merged into a single pen-based visual input modality, allowing designers to visually express refinement intent---such as through arrows, circles, or short handwritten labels---directly on the image. This design reflects the natural integration of scribbles and annotations observed in the formative study and supports intuitive, sketch-like interaction.

\textbf{Combination.}
The combined modality integrates text and visual inputs into a single multimodal request, enabling flexible and adaptive switching between and combination of the two modalities in line with \textbf{DG3}. Designers can alternate between or combine pen-based scribbles, handwritten notes, and typed text descriptions to support more precise and expressive refinement. 

Region selection for editing follows the interaction conventions of each modality: text uses a brush-based \textit{inpainting} tool, while visual and combination modalities allow direct marking of edit regions via pen-based input (e.g., circles, arrows).

\subsection{User Interface}
\begin{figure*}[t]
    \centering
    \includegraphics[width=\linewidth]{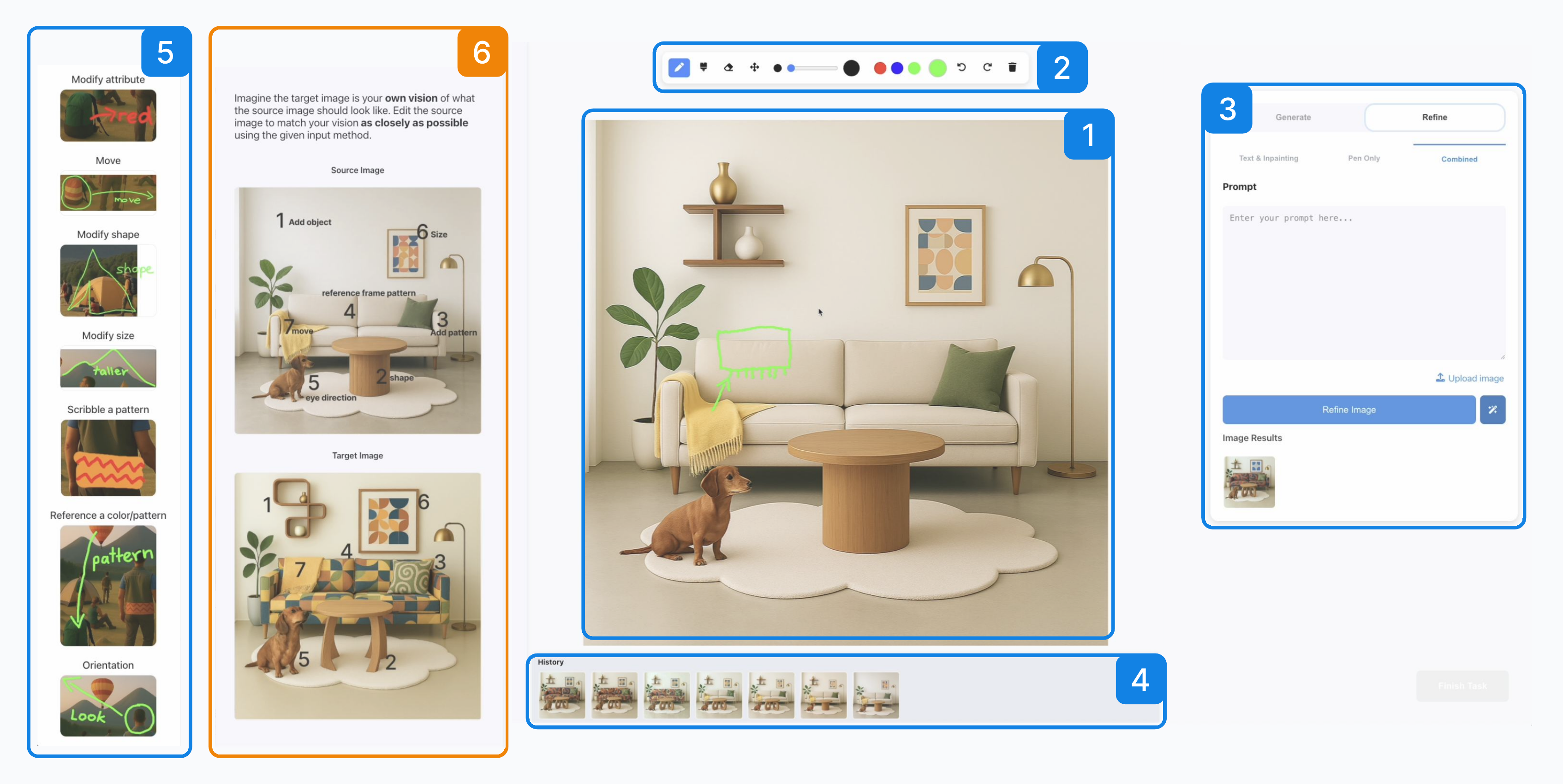} 
    \caption{User interface of the multimodal prompting prototype in the \textbf{Combination} modality. 
    (1) Central canvas supporting scribbles, annotations, and inpainting; 
    (2) Toolbar with drawing and editing tools (pen, inpainting, eraser, canvas dragging, stroke width, color selection, undo/redo, and clear canvas); 
    (3) Prompt panel for text input and submission; 
    (4) History panel displaying previous generations for iterative refinement; 
    (5) Visual prompt lexicon panel showing visual prompt examples; and 
    (6) Task panel displaying source and target images, visible only during closed-ended tasks to illustrate refinement goals.}
    \label{fig:user_interface_components}
    \Description{The figure shows the user interface of the multimodal visual prompting prototype in the combination modality, annotated with numbered regions. The main area is a central canvas (1) where users draw scribbles, add annotations, and apply inpainting directly on the image. Above the canvas is a toolbar (2) providing drawing and editing controls. A prompt panel (3) on the right supports text input and refinement submission. Below the canvas, a history panel (4) displays previously generated images to support iterative refinement. On the left, a visual prompt lexicon panel (5) presents example visual prompts for reference. A task panel (6) shows the source and target images for closed-ended tasks, clarifying refinement goals.}
\end{figure*}

The interface maintains a consistent layout across input modalities to isolate the effect of the modality from other interface components. As shown in \autoref{fig:user_interface_components}, it consists of:
(1) a central canvas displaying the editable GenAI-generated image;
(2) a top toolbar providing drawing and inpainting tools (pen, inpainting, eraser, stroke width, color, undo/redo, clear canvas, and canvas movement);
(3) a right panel for text input and refinement submission;
(4) a bottom history strip showing previous generations for incremental refinement \textbf{(DG4)};
(5) a left panel featuring a visual prompt lexicon that illustrates example scribbles and annotations to support interpretation of visual inputs \textbf{(DG5)}; and
(6) a task panel displaying the source and target images, visible only during closed-ended tasks. Irrelevant tools are hidden in each modality (e.g., pen tools in \textbf{Text-only} or text fields in \textbf{Visual-only} are not shown) to reduce cognitive load and interface clutter.

\subsection{Implementation}
The prototype\footnote{Prototype code, task assets, and configuration details are available at \url{https://github.com/parkhyerim/enhancing-genai-image-refinement}.} was implemented as a web-based application with a React front end and a Python (FastAPI) backend. Stylus input was supported via a Huion pen tablet, while keyboard and mouse interactions were used for text input and inpainting. Two GenAI image models were integrated based on pilot testing with two design students who compared output quality, latency, and stability across different task types. \textit{FLUX.1 Kontext Pro (FLUX.1)} handled most refinement tasks efficiently, particularly contextual spatial edits such as object movement and orientation changes, with average response times of 10--15 seconds. However, repeated iterations occasionally altered unrelated regions or slightly degraded overall image quality.
\textit{GPT-Image-1} was primarily used for object-addition tasks, producing higher-quality, more coherent results across input modalities. Although slower (around 45 seconds, approximately three to four times longer than \textit{FLUX.1}), pilot participants preferred its visual fidelity. To balance speed and quality, \textit{FLUX.1} was used for most refinements, while \textit{GPT-Image-1} was selectively employed for adding new objects or regenerating degraded outputs.

\begin{figure*}[t]
    \centering
    \includegraphics[width=\linewidth]{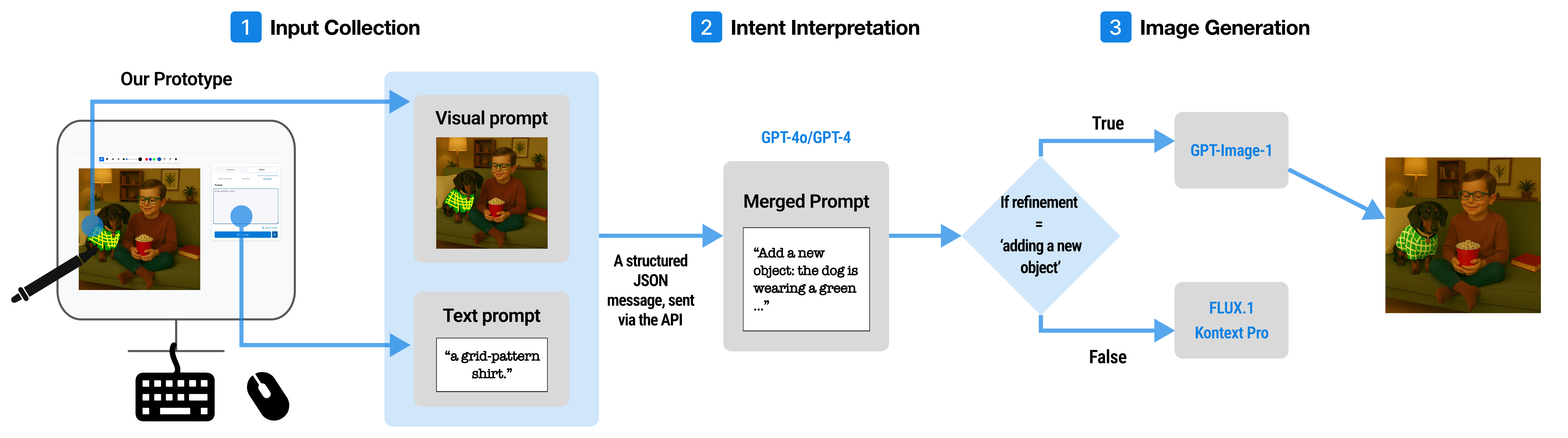}
    \caption{System pipeline of the prototype for image refinement. The process includes: (1) collecting text, visual prompts, and inpainting masks (\textbf{Input Collection}); (2) interpreting and merging them into a structured intent using \textit{GPT-4} or \textit{GPT-4o} (\textbf{Intent Interpretation}); and (3) generating the refined image with either \textit{GPT-Image-1} for object additions or \textit{FLUX.1 Kontext Pro} for contextual refinements (\textbf{Image Generation}).}
    \label{fig:pipeline}
    \Description{The figure illustrates the system pipeline for image refinement, organized into three sequential stages. In the first stage, input collection, users provide text prompts, visual prompts such as scribbles or annotations, and optional inpainting masks through the prototype interface. In the second stage, intent interpretation, the system merges these inputs into a structured representation of user intent. A decision step determines whether the refinement involves adding a new object. In the third stage, image generation, the system routes the structured intent to different image generation models depending on the refinement type, producing the final refined image. The figure shows how multimodal inputs are transformed and processed through the pipeline.}
\end{figure*}

\textbf{Processing pipeline.}
Inputs from each modality followed a standardized three-step process (\autoref{fig:pipeline}):  
(1) \textit{Input collection}---users provide text prompts, visual prompts, and inpainting masks (region selection);  
(2) \textit{Intent interpretation}---these inputs are converted into a structured JSON message and processed by \textit{GPT-4} or \textit{GPT-4o} to extract a structured textual description of the user's refinement intent; and  
(3) \textit{Image generation}---the interpreted intent is sent to the selected GenAI model: \textit{GPT-Image-1} for object additions and \textit{FLUX.1} for other refinement tasks.  
For \textbf{Visual-only} input, \textit{GPT-4} interprets annotations and scribbles, while for \textbf{Combination} input, \textit{GPT-4o} merges text, inpainting masks, and pen inputs into a unified representation.
To reduce misinterpretation from model variability or ambiguous shorthand input, intent interpretation was constrained to seven predefined refinement types (\textit{add, reshape, modify pattern, reference, reorient, resize, move}), defined through system prompts for structured control and consistent participant experience (Appendix~\ref{app:system_prompts}). The system automatically categorized user inputs into these types to guide image generation while maintaining model flexibility.

\section{User Study} 
\label{sec:userstudy}
This section details the user study, including the overview, participants, study design, measures, procedure, and analysis methods used to address the four research questions.

\subsection{Study Overview}
To evaluate how different input modalities support GenAI image refinement, we conducted a within-subjects study comparing \textbf{Text-only}, \textbf{Visual-only}, and \textbf{Combination} prompting using the prototype described in \autoref{sec:prototype}. The study examined how each modality affected the expression of refinement intent (\textbf{RQ1}); efficiency (task-completion time), workload, and user experience (\textbf{RQ2}); iterative refinement (\textbf{RQ3}); and how users adopted and switched between input modalities within the \textbf{Combination} condition (\textbf{RQ4}).

Based on insights from the formative study, we hypothesized that the \textbf{Visual-only} and \textbf{Combination} modalities would enable clearer intent expression than \textbf{Text-only} (\textbf{H1}); that \textbf{Visual-only} would improve efficiency and reduce workload, and \textbf{Combination} would enhance user experience (\textbf{H2}); that \textbf{Combination} would better support iterative refinement (\textbf{H3}); and that \textbf{Combination} would encourage flexible switching or integration of modalities depending on refinement type (\textbf{H4}).

\subsection{Participants}
Thirty participants (16 female, 14 male) participated in the study. Seventeen were professional designers and thirteen were design students, representing diverse domains, including industrial, UX/UI, graphic, architectural, and automotive interior design. Professional experience ranged from six months to 30 years. Participants' experience with GenAI image tools varied from daily to less than monthly; most (29/30) had used at least one system such as Midjourney, DALL-E, or Firefly. Twelve participants reported using a stylus at least occasionally in their workflow, while others primarily used mouse and keyboard input. Recruitment was conducted through company and university mailing lists, personal networks, and snowball sampling. Participants outside the collaborating company received non-monetary compensation valued at EUR 29. \autoref{tab:demo} summarizes participant demographics, design roles, years of design experience, GenAI image tools used, use frequency, and stylus use.

\begin{table*}[t]
    \centering
    \footnotesize
    \caption{Participant demographics ($N = 30$), including gender (G), design role, years of experience (YoE), GenAI image tools used, use frequency, and stylus use. Abbreviations: <Monthly = less than once a month; Monthly = once a month; 2--3$\times$/wk = two to three times per week.}
    \label{tab:demo}
    \Description{The table summarizes demographic information for 30 participants. Each row corresponds to one participant, and columns report gender, design role, years of experience, GenAI image tools used, frequency of GenAI use, and frequency of stylus use. The table provides an overview of participants' professional backgrounds and prior experience with GenAI image tools.}
    {\rowcolors{2}{gray!7}{white}
    \begin{tabularx}{\linewidth}{r c X c X c c}
        \toprule
        \textbf{ID} & \textbf{G} & \textbf{Design Role} & \textbf{YoE} & \textbf{GenAI Image Tools} & \textbf{GenAI Use} & \textbf{Stylus Use} \\
        \midrule
        1 & M & Creative Director & 16 & Midjourney, RunwayML, Kling AI & Daily & Weekly \\
        2 & M & Designer / Creative Technologist & 4 & ComfyUI, Flux & 2--3$\times$/wk & Monthly \\
        3 & M & Architect & 25 & None & 2--3$\times$/wk & No \\
        4 & M & Design Lead (Concept/UX/UI/Interior) & 26 & ChatGPT, Midjourney, RunwayML & Monthly & No \\
        5 & F & Architect & 30 & Midjourney, DALL-E & Monthly & No \\
        6 & F & Junior UX/Service Designer & 3.5 & DALL-E, Firefly & Weekly & No \\
        7 & M & UX/UI Developer & 5 & DALL-E, Firefly, Grok & Monthly & No \\
        8 & F & Graphic Designer / Communicator & 6 & Internal company tool & Monthly & <Monthly \\
        9 & M & UI/UX Designer (Early Phase) & 18 & Midjourney, DALL-E, Reve & Weekly & Weekly \\
        10 & F & Innovation Manager (Digital Products) & 10 & Firefly, Midjourney, DALL-E & Weekly & No \\
        11 & M & Contemporary Artist & 7 & Firefly & Monthly & Weekly \\
        12 & M & UX Designer & 16 & Midjourney, Stable Diffusion, ComfyUI & 2--3$\times$/wk & No \\
        13 & M & UX / Concept Designer & 8 & Midjourney, DALL-E, Vizcom, RunwayML, Firefly & Monthly & 2--3$\times$/wk \\
        14 & F & UX Designer & 15 & Midjourney, DALL-E & Monthly & No \\
        15 & F & Architecture Student (Master) & 10 & Sora & <Monthly & No \\
        16 & M & Exterior Designer & 3 & Midjourney, Vizcom & Weekly & Monthly \\
        17 & M & Industrial Design Student & 3 & Midjourney, DALL-E, Firefly, Vizcom, Krea, Recraft & Weekly & No \\
        18 & F & UX Design Student (Concept) & 2 & DALL-E, Firefly & 2--3$\times$/wk & No \\
        19 & F & Computer Science \& Design Student & 3.5 & Midjourney, DALL-E, Firefly, Stable Diffusion, Imagen, Flux & Monthly & Daily \\
        20 & F & UX Design Student (Work-Study Program) & 2 & Firefly & Weekly & No \\
        21 & F & Interior Designer & 17 & Midjourney, DALL-E & Weekly & No \\
        22 & F & UI/UX Design Student (Working Student) & 3 & DALL-E, Firefly & Weekly & No \\
        23 & M & Industrial Design Intern & 2.5 & ChatGPT, ComfyUI & 2--3$\times$/wk & <Monthly \\
        24 & M & Industrial Design Intern / Student & 4 & Firefly, Midjourney, DALL-E, ComfyUI & 2--3$\times$/wk & No \\
        25 & F & Communications Design Student / Intern & 1.5 & Midjourney, DALL-E & <Monthly & No \\
        26 & F & Interactive Media Student & 3 & Firefly & <Monthly & No \\
        27 & F & Art \& Multimedia Student & 2.5 & Midjourney, DALL-E & Weekly & Daily \\
        28 & F & Car Interior Designer & 5 & Midjourney, Krea & <Monthly & <Monthly \\
        29 & F & UI/UX Design Student & 2.5 & ComfyUI, Midjourney, DALL-E & 2--3$\times$/wk & Monthly \\
        30 & M & UX Design Student (Prototyping) & 0.5 & Stable Diffusion, DALL-E, GPT-Image-1 & 2--3$\times$/wk & No \\
        \bottomrule
    \end{tabularx}
    }
\end{table*}

\subsection{Study Design}
We used a within-subjects design to minimize individual variance across three input modalities:
\begin{enumerate}[leftmargin=1.5em]
    \item \textbf{Text-only (baseline):} keyboard input with inpainting.  
    \item \textbf{Visual-only:} pen-based scribbles and annotations using a tablet and stylus.  
    \item \textbf{Combination:} free use of text, visual input, or both. 
\end{enumerate}

Input modality order was counterbalanced across six Latin square sequences. Participants completed two phases: a structured, closed-ended phase and an exploratory, open-ended phase. The closed-ended phase ensured comparability across modalities, while the open-ended phase captured more naturalistic refinement behaviors. Details of both phases are provided in \autoref{sec:procedure}. 
As the study was conducted in English, we recruited participants who were comfortable writing in English. Text entered through typed prompts and annotations was encouraged to be in English to ensure consistency in model interpretation and analysis. However, participants were permitted to use their native language when they could not readily find an appropriate English term.

\subsection{Measures}
We collected both quantitative and qualitative data (full questionnaires and scales are provided in Appendix~\ref{app:questionnaire}):
\begin{enumerate}[leftmargin=*]
    \item \textbf{Subjective measures:} five NASA-TLX subscales (mental demand, temporal demand, performance, effort, and frustration), the UEQ-S (pragmatic and hedonic qualities), and two custom Likert-scale items evaluating the expression of refinement intent and support for iterative refinement.
    \item \textbf{Behavioral measures:} task completion time, number of iterations, input modality usage logs, and generated images.
    \item \textbf{Qualitative data:} think-aloud comments and post-study interviews, which captured participants' preferences, strategies, and perceived strengths and limitations of each modality.
\end{enumerate}

\textbf{Mapping to RQs.}
The study's measures and data sources were aligned with the four research questions as follows:
\begin{enumerate}[leftmargin=*]
    \item \textbf{RQ1:} Expression of refinement intent item and input--output image analysis.
    \item \textbf{RQ2:} Task completion time, NASA-TLX, and UEQ-S.
    \item \textbf{RQ3:} Iterative refinement item and qualitative interview analysis.
    \item \textbf{RQ4:} Multimodal usage patterns from input modality usage logs and post-study interviews.
\end{enumerate}

\begin{figure*}[t]
    \centering
    \includegraphics[width=\linewidth]{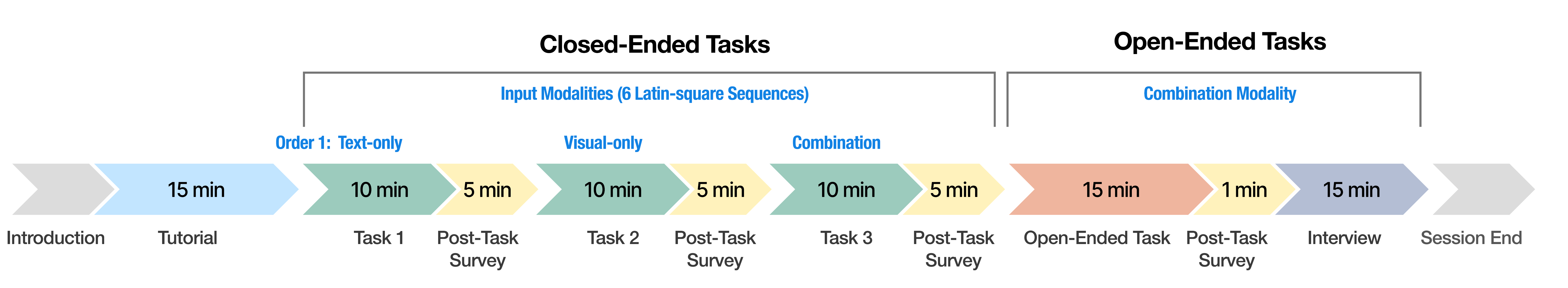}    
    \caption{Overview of the user study procedure. Each session (approximately 90 minutes) included an introduction, tutorial, three closed-ended tasks with alternating input modalities (\textbf{Text-only}, \textbf{Visual-only}, and \textbf{Combination}), and an open-ended task conducted exclusively in the \textbf{Combination} modality. Surveys followed each task, and the session concluded with an interview. The closed-ended phase was counterbalanced using a Latin square design with six sequences (example illustrated: \textbf{Order 1}).}
    \label{fig:study_procedure}
    \Description{The figure shows a timeline of a single user study session lasting approximately 90 minutes. The session begins with an introduction and tutorial, followed by three closed-ended tasks. Each closed-ended task uses a different input modality---text-only, visual-only, or combination---and is followed by a short post-task survey. The order of these three tasks is counterbalanced across participants using a Latin square design. After the closed-ended phase, participants complete an open-ended task conducted in the combination modality, followed by a post-task survey and a concluding interview.}
\end{figure*}

\subsection{Procedure}
\label{sec:procedure}
Each in-person session lasted approximately 90 minutes and consisted of five phases (\autoref{fig:study_procedure}). All participants provided informed consent prior to participation, and the study was conducted in accordance with institutional ethics requirements.

\paragraph{Introduction.}
Participants were briefed on the study objectives and procedure, and provided informed consent.

\paragraph{Tutorial and Practice.}
Participants were introduced to the interface, input modalities, and their corresponding devices, task types, and the visual prompt lexicon. They then practiced the \textbf{Text-only}, \textbf{Visual-only}, and \textbf{Combination} conditions, in that order, completing up to seven sample refinements covering the seven main task types. The planned 15-minute tutorial was flexibly adjusted based on participants' confidence and the experimenters' observations, and participants proceeded when they felt ready. The same image set and tasks were used across all modalities.

\paragraph{Closed-Ended Tasks.}

\begin{figure*}[t] 
    \centering
    \includegraphics[width=.95\linewidth]{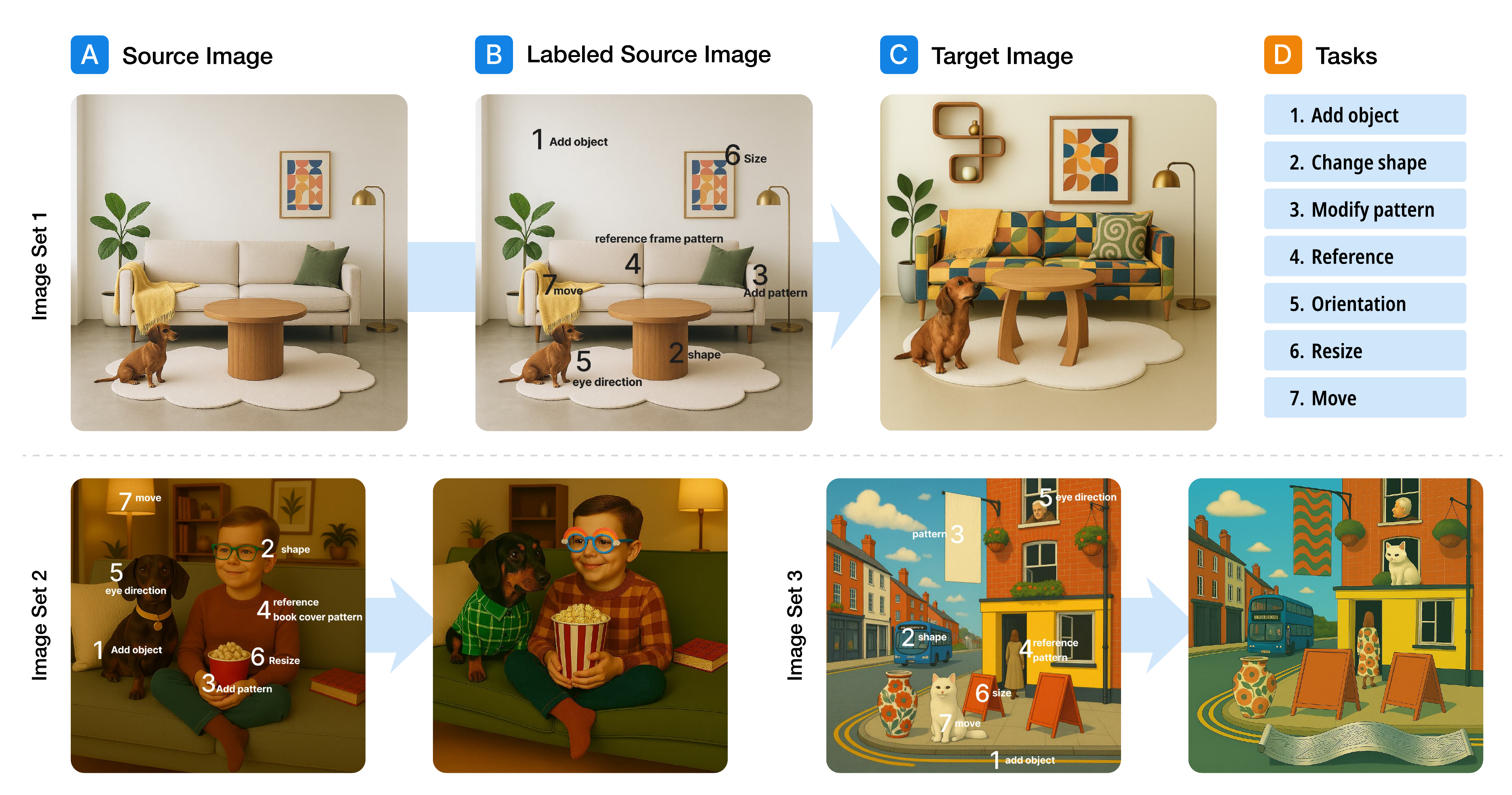} 
     \caption{Closed-ended tasks: (upper) three image sets were provided, each used with a different input modality depending on the assigned Latin-square group. Each set included (A) a source image, (B) the same image annotated with seven refinement tasks, (C) a corresponding target image, and (D) an overview of the seven task types. (lower) Example labeled source and target images from Sets~2 and~3. Image sets were generated in advance with \textit{DALL-E 3} and \textit{FLUX.1} to ensure all participants worked with identical materials.}
    \label{fig:close_ended}
    \Description{The figure illustrates the closed-ended task materials used in the study. The upper row shows the structure of an image set, consisting of a source image (A), the same image annotated with seven numbered refinement tasks (B), a corresponding target image showing the intended outcome (C), and a list of the seven task types (D). The lower row presents example labeled source and target images from two different image sets. Together, the panels show how predefined refinement tasks were specified and paired with target outcomes for controlled evaluation across input modalities.}
\end{figure*}

The closed-ended phase ensured consistent refinement goals across participants, enabling systematic comparison across conditions. They completed all three input modalities in a counterbalanced order (see Appendix~\ref{app:study_order}). Each image set included a source image, an annotated version with seven task labels, and a target image. Participants refined source images to match target images as closely as possible across seven predefined refinement tasks adapted and extended from the formative study (\autoref{fig:close_ended}): (T1) \textit{adding an object}, (T2) \textit{changing an object's shape}, (T3) \textit{modifying a pattern}, (T4) \textit{referencing an attribute from another object}, (T5) \textit{changing orientation}, (T6) \textit{resizing an object}, and (T7) \textit{moving an object}.  
To minimize workflow disruption, a task panel displaying the source image with task labels and the corresponding target image was integrated into the interface (to the left of the working canvas), allowing participants to reference tasks without using external displays (\autoref{fig:user_interface_components}). The interface layout remained consistent across all modalities. Participants were instructed to complete all tasks once using the respective input modality. If needed, they could retry each task up to three times to make further adjustments. This procedure enabled evaluation of each modality's effectiveness on the first attempt and across multiple adjustment cycles, while keeping the overall session duration within a reasonable limit. Task completion time was measured from the first input event (typing, drawing, or writing) to the click on the submit button (``Refine Image''), thereby maintaining a consistent interaction flow across modalities. All inputs and outputs were logged. While the system generated images, participants were encouraged to think aloud---explaining their reasoning and expectations. After completing each input modality condition, they completed the NASA-TLX, UEQ-S, and the refinement intent questionnaire (see Appendix~\ref{app:questionnaire}). This phase specifically addressed \textbf{RQ1} (expressing refinement intent), \textbf{RQ2} (efficiency, workload, and user experience), and \textbf{RQ3} (iterative refinement).

\paragraph{Open-Ended Tasks.}
This phase aimed to capture more natural refinement behavior and creative exploration, simulating real-world GenAI workflows. Participants first generated an image of their choice using a text prompt, then refined it freely using any input modality. The entire phase lasted approximately 15 minutes. No target images were provided; participants had full control over subject matter, refinements, and choice of modality. For the initial generation, text prompting was used because the \textit{dall-e-3} API produced more consistent results for text inputs, and the study focused on refinement rather than image generation from scratch.
The open-ended task allowed participants to refine self-generated images without predefined goals, revealing intrinsic motivation, novel strategies, and potential refinement tasks beyond the seven closed-ended ones. It also examined how participants naturally combined or switched input modalities. Refinements could be applied \textbf{iteratively and incrementally} over multiple attempts. This phase specifically addressed \textbf{RQ3} (iterative refinement) and \textbf{RQ4} (hybrid use and strategies). All inputs and outputs were logged, followed by a short questionnaire on iterative refinement (Appendix~\ref{app:iterative-refinement_questions}).

\paragraph{Post-Study Interview.}
A semi-structured interview (about 10 minutes) captured participants' preferences, strategies, and perceived advantages or limitations across input modalities. Before the interview, all participant-generated input and output images were displayed on the canvas to facilitate recall and discussion.

\paragraph{Experimental Setup.}
Sessions were conducted in a controlled laboratory environment with a MacBook Pro running the prototype, an external monitor, a Huion pen tablet, and a keyboard/mouse. Relevant input devices were enabled according to the active modality. All interaction events (text prompts, scribbles, annotations, and submission clicks) were logged, with screen and audio recordings captured during task performance and interviews.

\subsection{Data Analysis}
\subsubsection{Quantitative Analysis} 
Data normality was assessed using the Shapiro--Wilk test; as at least one condition violated this assumption for each scale, non-parametric tests were used. Overall differences between the three input modalities were examined with the Friedman test, followed by pairwise Wilcoxon signed-rank tests with Holm correction. We report $p < .05$ (two-tailed), effect sizes (Kendall's $W$ for Friedman tests and $r = Z / \sqrt{N}$ for Wilcoxon tests), and descriptive statistics (see Appendix~\ref{app:friedman_results}).  
Dependent variables included NASA-TLX and UEQ-S (\textbf{RQ2}) and two study-specific scales---\textit{Expression of Refinement Intent} (\textbf{RQ1}) and \textit{Iterative Refinement} (\textbf{RQ3}) scales---both showing high reliability ($\alpha = .88$ and $\alpha = .92$).  Efficiency, workload, and user experience (\textbf{RQ2}) were additionally assessed via task completion time. All analyses and visualizations were conducted in Python.

\subsubsection{Qualitative Analysis}
Interview and think-aloud data were analyzed using thematic analysis~\cite{cooper_thematic_2012, braun_one_2021}. Two authors performed open coding, iteratively grouping codes into subthemes and organizing these into overarching themes aligned with the research questions. The coding framework was refined through regular discussions to reach consensus and ensure consistency. \textbf{RQ4} (hybrid use and strategies) was mainly addressed through qualitative analysis, supported by descriptive analysis of multimodal usage logs.

\section{Results}
\label{sec:results}
Results are organized by research question: \textbf{RQ1} (expressing refinement intent), \textbf{RQ2} (efficiency, workload, and user experience), \textbf{RQ3} (iterative refinement), and \textbf{RQ4} (hybrid use and strategies). Each subsection integrates quantitative and qualitative findings. 
Detailed statistics and additional figures are provided in Appendix~\ref{app:quant_analysis}.

\subsection{RQ1: Expressing Refinement Intent}
Ratings on the Expression of Refinement Intent scale differed significantly across input modalities ($\chi^2(2) = 29.04$, $p < .001$, $W = .48$). Both \textbf{Visual-only} and \textbf{Combination} modalities were rated higher than \textbf{Text-only} across most refinement types, with particularly strong effects for spatial relations and size or scale adjustments (both $p < .001$, $|r| > .7$). \textbf{Combination} achieved the highest overall scores, outperforming \textbf{Text-only} across all items and exceeding \textbf{Visual-only} for local edits and visual features (see \autoref{fig:expressing_intent}; detailed results in Appendix~\ref{app:expressing_intent_results}). 

\begin{figure}[t]
    \centering
    \includegraphics[width=\linewidth]{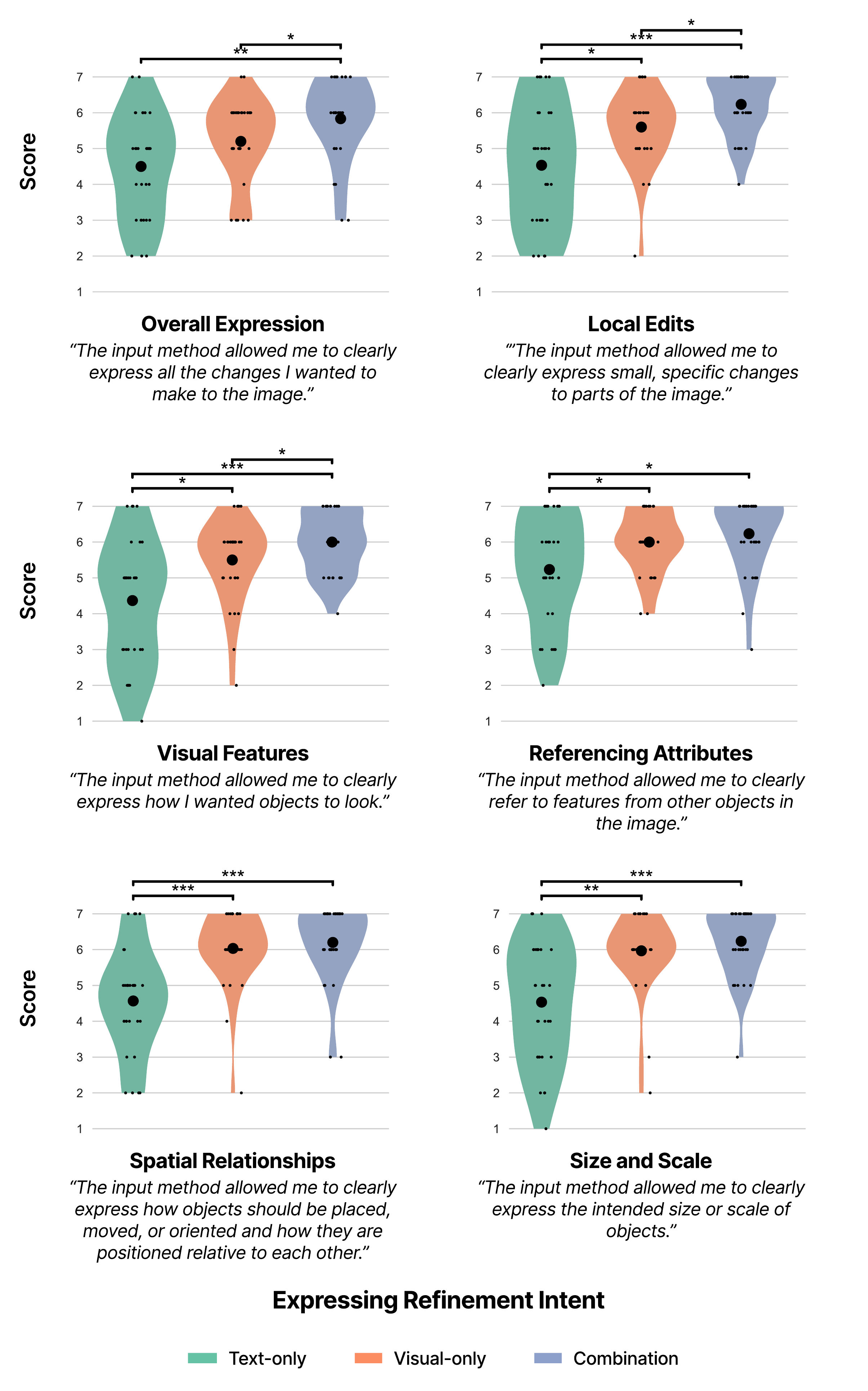}
    \caption{Expression of refinement intent ratings across modalities. \textbf{Visual-only} and \textbf{Combination} outperformed \textbf{Text-only}, with the \textbf{Combination} modality rated highest overall.} 
    \label{fig:expressing_intent}
    \Description{The figure presents six violin plots comparing participant ratings of how well different input modalities support expressing refinement intent. Each subplot corresponds to a refinement aspect: overall expression, local edits, visual features, referencing attributes, spatial relationships, and size and scale. Within each subplot, distributions are shown for three modalities: text-only, visual-only, and combination, with individual data points overlaid and central tendency indicated.}
\end{figure}

Qualitative findings aligned with these results. Participants emphasized that visual input provided clarity and speed in concrete spatial adjustments, orientation changes, and object relations. Arrows, circles, and scribbles were described as faster and less ambiguous than text when linking source and target elements:  
\textit{``The look direction\ldots describing it is harder than just drawing an arrow (\ldots) these are very practical using scribbles''} (P13). Such pen-based input also improved precision and speed in complex scenes:  
\textit{``If it was just text I'd have to specify the middle window (\ldots) the arrow definitely makes more sense''} (P23). However, several participants noted limitations, including handwriting recognition errors (P4) and the effort or imprecision involved in drawing detailed attributes such as color or texture. In contrast, text input was preferred for global or conceptual refinements (e.g., ``sky,'' ``futuristic'') and for naming specific or well-known objects (e.g., ``Eiffel Tower,'' ``double-decker bus'').  
As P25 explained, \textit{``If I wanted to have a golden retriever, I would write it down because the AI knows what it looks like better than I can sketch.''}  
However, text was often less effective for expressing spatial or shape-related relations, and several non-native English speakers mentioned vocabulary barriers (P13, P25).  
Combined inputs allowed participants to merge these strengths---using sketches for spatial intent and short text for detail---effectively combining spatial clarity with semantic precision: \textit{``Using scribbles and then text, you get the best of both worlds''} (P1).

\subsection{RQ2: Efficiency, Workload, and User Experience}
\subsubsection{Efficiency (Task Completion Time)}
Task completion times differed significantly across input modalities for spatially oriented refinements (Tasks~4--7: referencing, orientation changes, resizing, and moving) ($\chi^2(2) = 18.41$, $p < .001$, $W = .32$), but not for conceptual edits (Tasks~1--3: adding, shape changes, and pattern addition). Post-hoc comparisons showed that \textbf{Visual-only} was consistently faster than \textbf{Text-only} ($p < .01$). Orientation changes were completed in roughly half the time with \textbf{Visual-only} ($M = 16.9\,\text{s}$) compared to \textbf{Text-only} ($M = 30.3\,\text{s}$). No significant differences were observed between \textbf{Visual-only} and \textbf{Combination}, indicating that adding text did not increase task completion time. Although Tasks~1--3 showed no significant differences, \textbf{Text-only} refinements often took longer due to prompt reformulation, while \textbf{Visual-only} refinements involved more extended sketching. Removing outliers confirmed the same pattern: \textbf{Visual-only} and \textbf{Combination} were faster for spatially demanding edits (Appendix~\ref{app:task_completion_time}).  

Qualitative feedback supported these results. Participants described visual input as faster and more natural for spatial refinements such as resizing, moving, and changing orientation compared to lengthy text descriptions:  
\textit{``It's easier with scribbles; I can just draw an arrow without thinking (\ldots) position is hard to describe in text''} (P28).  
In comparison, text prompts were preferred for broad or global edits such as lighting, material, or overall color tone, where drawing would be more time-consuming, although no significant time advantage was observed in the quantitative results:  
\textit{``When I say there should be a sky (\ldots) why should I paint a sky? It's a huge area, so you would use a text prompt''} (P15).  
Combined inputs did not increase task completion time and were often used redundantly to ensure clarity: \textit{``I always think it couldn't be worse if I write something to specify it''} (P29).

\subsubsection{Workload and Effort (NASA-TLX)}

\begin{figure}[t]
    \centering
    \includegraphics[width=\linewidth]{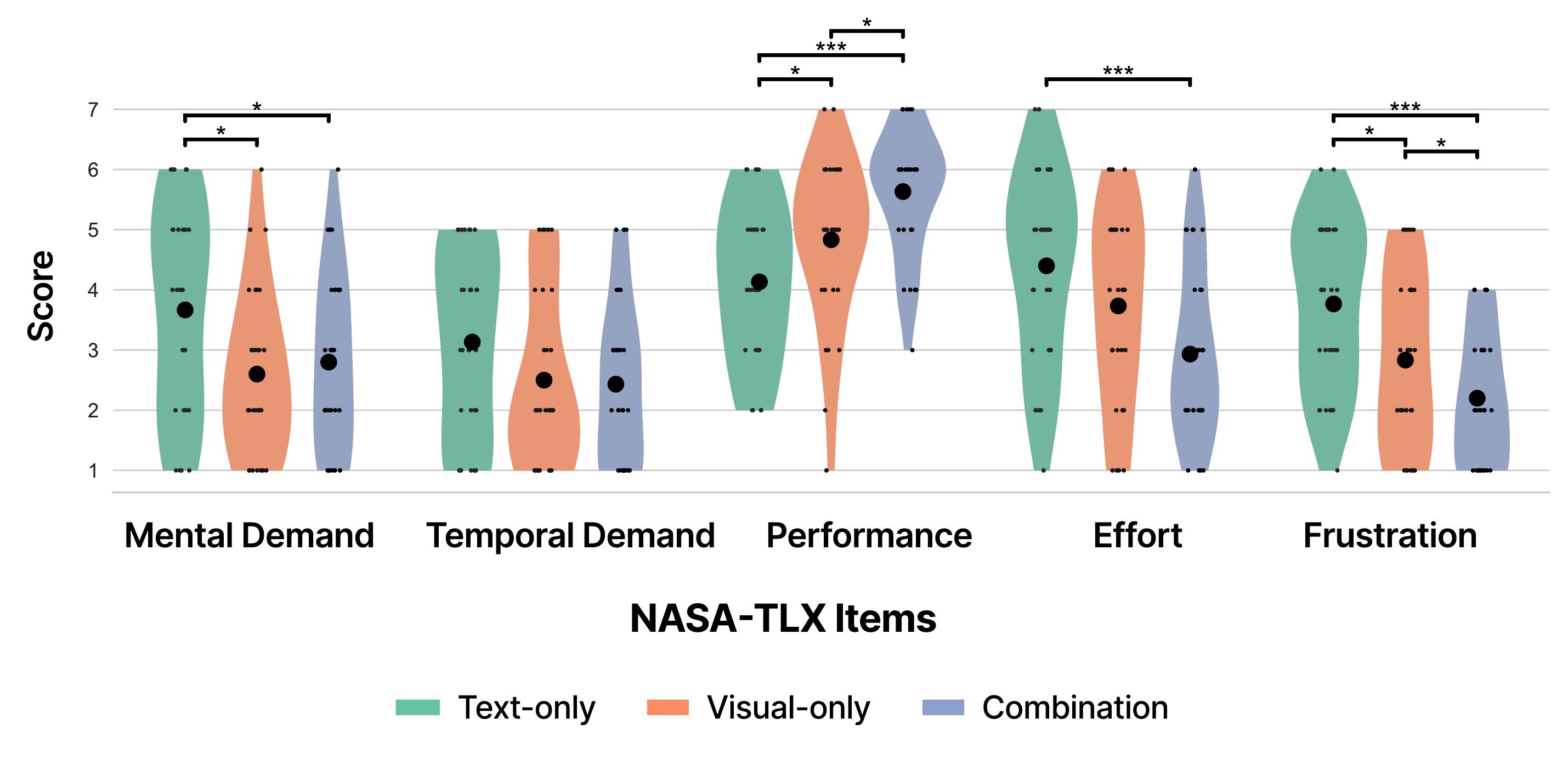}
    \caption{NASA-TLX results across modalities. Lower scores indicate less workload; performance is reversed. \textbf{Visual-only} and \textbf{Combination} lowered workload and frustration relative to \textbf{Text-only}.}
    \label{fig:nasa_tlx}
    \Description{The figure shows violin plots of NASA-TLX ratings across input modalities. Each subplot corresponds to a workload dimension, including mental demand, temporal demand, performance, effort, and frustration, with side-by-side violins for text-only, visual-only, and combination modalities. The vertical axis shows Likert-scale scores, with lower values indicating lower workload, except for performance which is reversed. Violin width reflects response distributions, and brackets with asterisks indicate statistically significant differences between modalities.}
\end{figure}

A Friedman test revealed significant differences across all five NASA-TLX dimensions ($p < .05$). Mean scores showed that \textbf{Visual-only} reduced mental demand, temporal demand, effort, and frustration compared to \textbf{Text-only}, while performance was rated higher (see Appendix~\ref{app:nasa_tlx_tb}). Post-hoc comparisons showed large effects for mental demand and frustration ($p < .01$). \textbf{Combination} provided additional benefits, yielding higher performance and lower frustration than \textbf{Visual-only}, while temporal demand did not differ significantly across modalities (\autoref{fig:nasa_tlx}).

Qualitative insights echoed these results. Participants described visual input as intuitive and confidence-building: \textit{``When you draw something, it gives you more confidence that the result will match your intentions''} (P12). They also found it enjoyable: \textit{``I just have fun if I can directly draw on the image''} (P27). However, uncertainty about the system's handwriting recognition, as well as the effort required for detailed drawings, sometimes increased mental demand and frustration, making typing feel more reliable and less taxing (P12, P22, P23). Some participants found text input less demanding due to familiarity with text-based GenAI tools: \textit{``With just using this text box, I'm very used to it (\ldots) typing is more user-friendly''} (P23). Combined inputs reduced uncertainty but occasionally disrupted workflow due to device switching, as one participant noted: \textit{``Switching tools costs time and changes your mental model''} (P4).

\subsubsection{User Experience (UEQ-S)}
Perceived user experience was assessed using the UEQ-S (Appendix~\ref{app:ueq-s}), with responses transformed from the original 1--7 scale to the standard –3 to 3 range. A Friedman test revealed significant differences across all UEQ-S dimensions ($p < .001$). Mean ratings indicated that \textbf{Visual-only} and \textbf{Combination} yielded higher pragmatic quality (e.g., efficiency, ease of use), hedonic quality (e.g., interesting, inventive), and overall scores compared to \textbf{Text-only}. Post-hoc comparisons confirmed large effects ($p < .001$): both modalities outperformed \textbf{Text-only}, with no difference between them for hedonic quality. For pragmatic quality, all pairwise differences were significant ($p < .05$), with \textbf{Combination} rated highest overall (\autoref{fig:ueq_plot}; detailed results in Appendix~\ref{app:ueq-s_results}).

\begin{figure}[t]
    \centering
    \includegraphics[width=.95\linewidth]{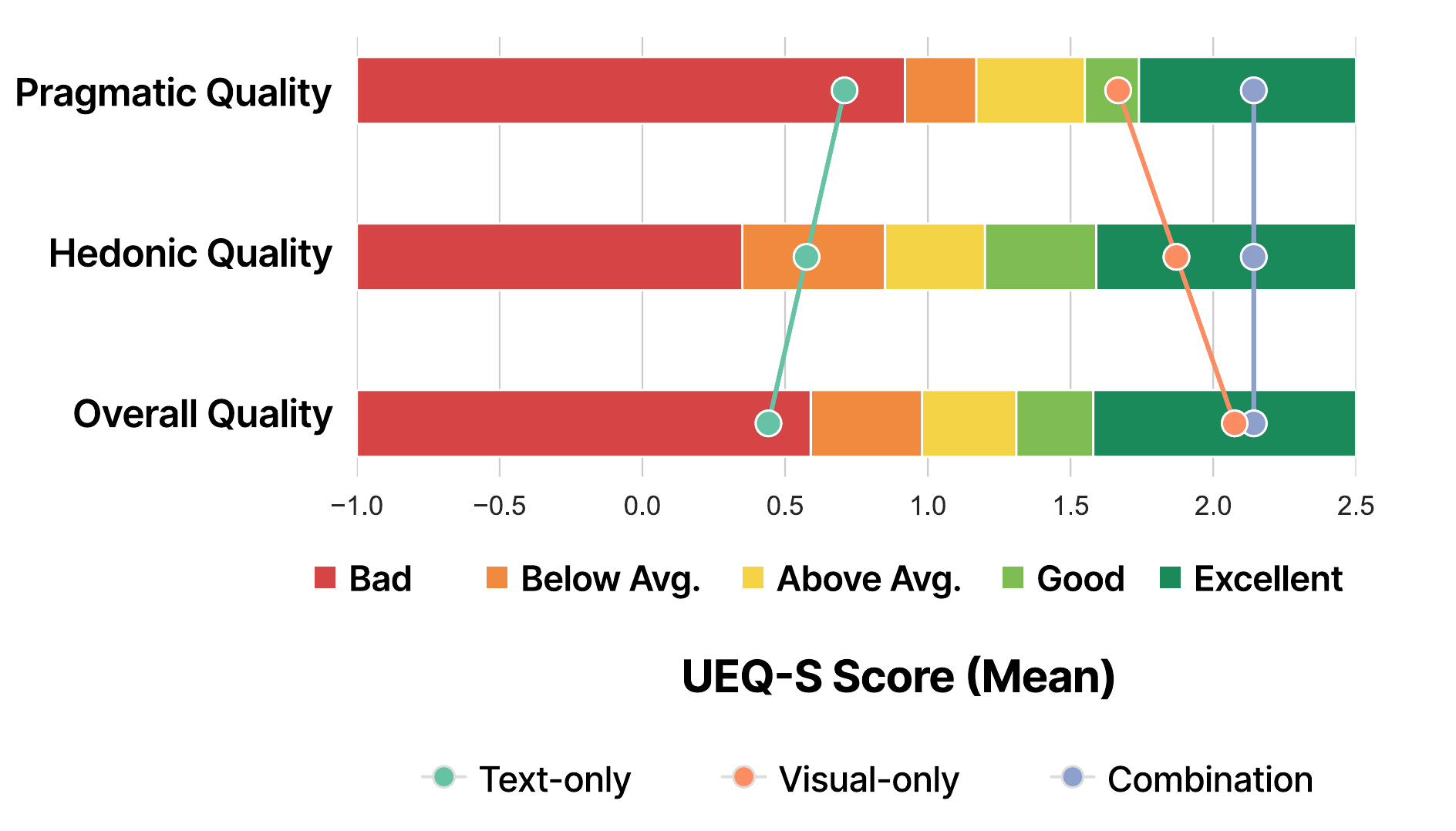}
    \caption{UEQ-S results across modalities. \textbf{Visual-only} and \textbf{Combination} scored higher than \textbf{Text-only} on pragmatic and hedonic quality, with the \textbf{Combination} modality rated highest overall.}
    \label{fig:ueq_plot}
    \Description{The figure shows mean UEQ-S scores across input modalities for pragmatic quality, hedonic quality, and overall quality. Each row represents a quality dimension, plotted against a horizontal scale ranging from bad to excellent. Colored markers indicate mean scores for text-only, visual-only, and combination modalities, with lines connecting modality scores within each dimension. The figure allows comparison of perceived user experience quality across modalities.}
\end{figure}

Participants described visual inputs as natural and enjoyable, particularly for those with drawing-oriented workflows or design backgrounds: \textit{``I'm an architect; I always like to draw everything''} (P3). Text input felt familiar and convenient for participants accustomed to typing or text-based GenAI tools, though several found it limiting in vocabulary and expressiveness. \textit{``It's really hard to describe a specific texture (\ldots) I cannot describe it in my language''} (P13). Combined input was consistently viewed as the most flexible and powerful, enabling users to refine ideas quickly while retaining precision through text (P1, P17, P28).

\subsection{RQ3: Iterative Refinement}
\begin{figure}[t]
    \centering
    \includegraphics[width=\linewidth]{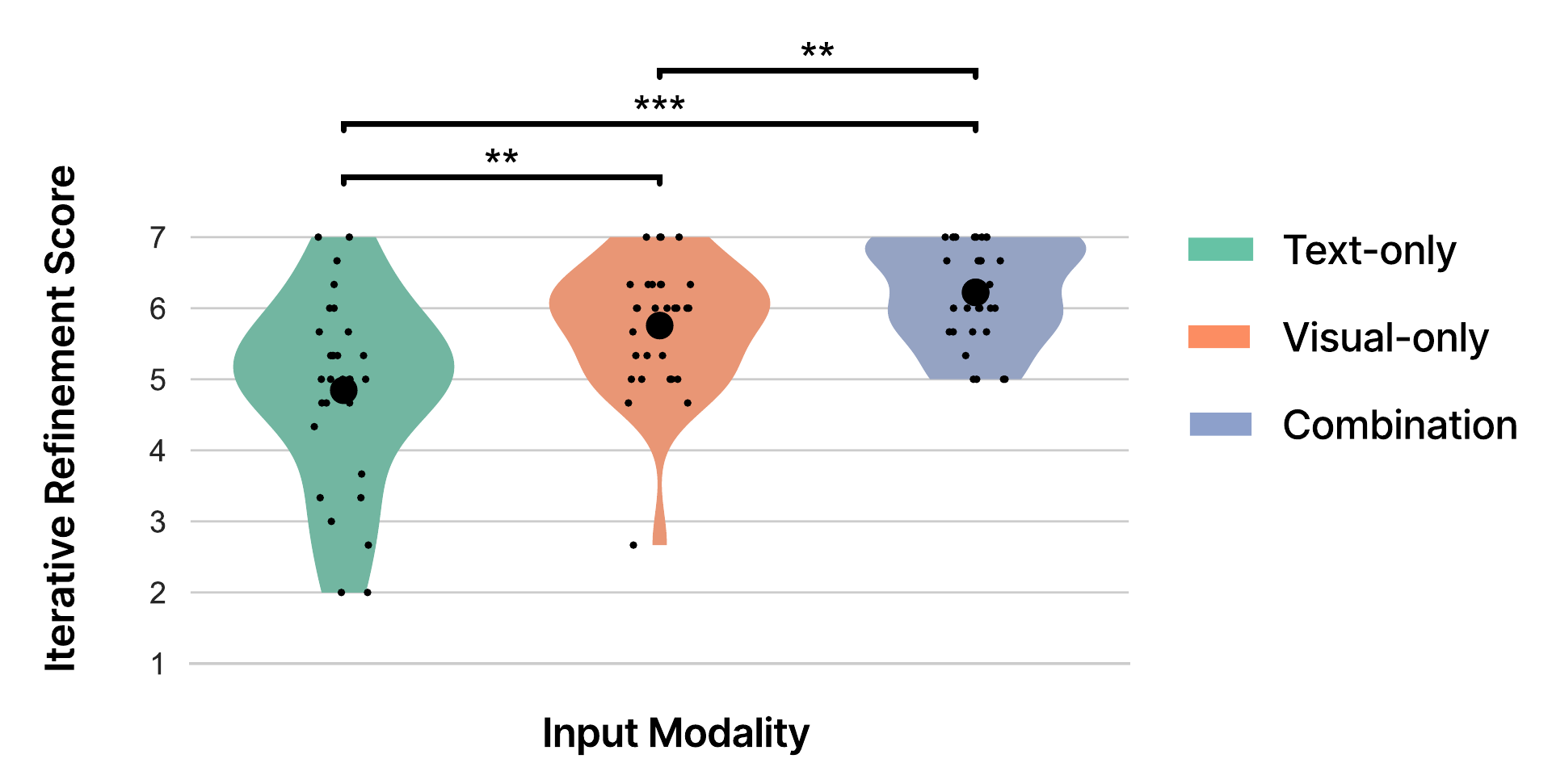}
    \caption{Iterative refinement ratings across modalities. \textbf{Combination} and \textbf{Visual-only} supported iterative refinement more effectively than \textbf{Text-only}.}
    \label{fig:iterative_plot}
    \Description{The figure shows ratings of support for iterative refinement across input modalities. Distributions are plotted for text-only, visual-only, and combination modalities along a Likert-scale vertical axis. The figure enables comparison of how different modalities support repeated refinement during interaction.}
\end{figure}

Ratings on the Iterative Refinement scale differed significantly across input modalities ($p < .001$). \textbf{Combination} scored highest ($M = 6.2$, $SD = 0.7$), followed by \textbf{Visual-only} ($M = 5.8$, $SD = 0.9$) and \textbf{Text-only} ($M = 4.8$, $SD = 1.3$). Post-hoc tests confirmed all pairwise differences ($p < .001$), indicating that multimodal and visual inputs supported iterative refinement more effectively than text-based input alone (\autoref{fig:iterative_plot}).  

Qualitative findings showed that visual input supported iterative refinement by enabling rapid, direct, and incremental adjustments without requiring users to translate ideas into text, thereby preserving flow and confidence. As one participant noted, \textit{``When you draw something, it gives you more confidence that the result will match your intentions''} (P12). Participants often began with rough sketches to externalize ideas and explore possibilities, then progressively refined spatial form and relationships with more detailed marks. This progression illustrates how visual ambiguity can be productive early in iteration--supporting reinterpretation and adjustment--before becoming more precise as intent stabilizes, consistent with prior accounts of sketch-based exploration~\cite{gross_ambiguous_1996}.
When using combined inputs, many described a \textit{visual-first workflow}: they started with quick scribbles to establish spatial intent and then added text to clarify semantic details or fine-grained constraints. This sequencing supported stepwise and controlled iteration because users could build on visible visual context while maintaining their refinement intentions across iterations.
By contrast, text input was preferred for conceptual or global variations (e.g., lighting, materials, tones) and for generating alternative ideas by leveraging GenAI's creative capabilities. However, text was less effective for local refinements, as it often required repeated trial--and--error prompting to achieve precision. Participants experienced this process as disruptive and frustrating rather than creatively iterative. At the same time, visual input reduced ambiguity in spatial interpretation and enabled participants to ``draw what they mean'' (P3, P12), resulting in faster and more predictable refinement across iterations.

\subsection{RQ4: Hybrid Use and Strategies in the Combination Condition}
To examine how participants used and combined input modalities in the \textbf{Combination} condition---where text and visual input were available but could be used independently or together---we analyzed logged inputs alongside interview data. Two primary strategies emerged: \textbf{complementary use}, in which visual and text inputs conveyed different aspects of the refinement intent, and \textbf{redundant use}, where both modalities expressed the same information to increase clarity and confidence.
\textbf{Complementary use} occurred when scribbles or annotations externalized spatial or compositional intent (e.g., object position, orientation), while text specified semantic or stylistic details (e.g., material, color, style):  
\textit{``Using scribbles to draw shapes that are hard to describe with text and then using text (\ldots)''} (P1).  
\textbf{Redundant use} involved repeating information across modalities to confirm or disambiguate rough sketches and counter uncertainty about system interpretation:  
\textit{``I always think it couldn't be worse if I write something to specify it''} (P29); \textit{``I just add both (\ldots) to make sure it's understood''} (P20).

\begin{figure}[t]
    \centering
    \includegraphics[width=\linewidth]{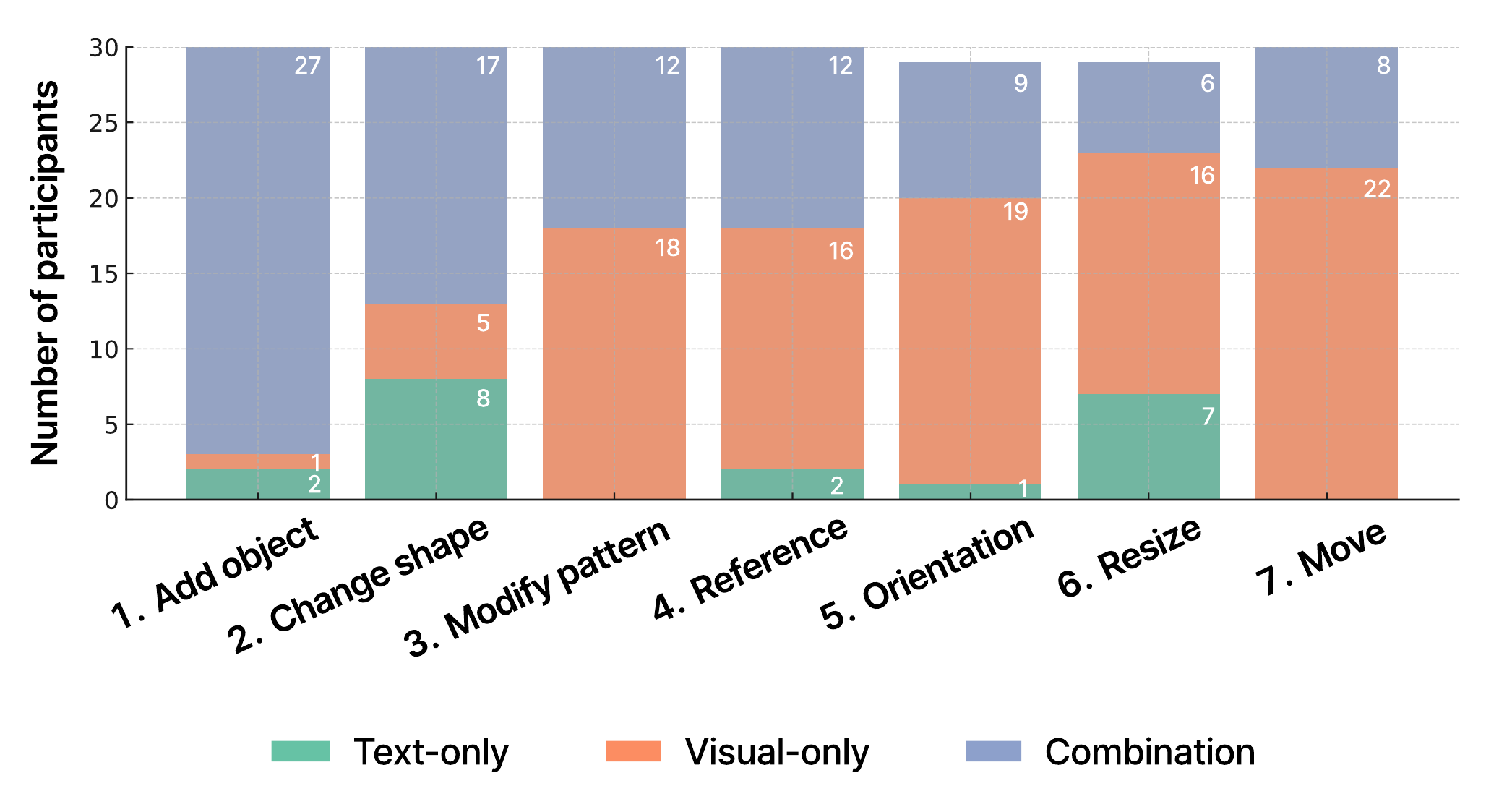}
    \caption{Input strategies (text, visual, and combined input) across seven refinement tasks in the \textbf{Combination} condition during the closed-ended phase. Participants had access to both text and visual input, which could be used independently or in combination. The figure shows modality use in the first iteration of each task, illustrating task-dependent selection: combined input was used most often for adding new content (Task~1), whereas visual input was used primarily for pattern modification (Task~3) and spatial adjustments (Tasks~4--7). Tasks~5--6 include $N = 29$ participants; all other tasks include $N = 30$. Per-participant distributions are provided in Appendix~\ref{app:strategies_results}.}
    \label{fig:input_strategy_summary}
    \Description{The figure shows stacked bar charts summarizing input strategies used across seven refinement tasks during the closed-ended phase. Each bar corresponds to a task and is divided into segments representing the number of participants who used text-only, visual-only, or combined input in the first iteration. The vertical axis indicates participant counts, allowing comparison of how input strategies were distributed across different task types.}
\end{figure}

\paragraph{Task-specific patterns.} 
Task type strongly influenced modality choice (\autoref{fig:input_strategy_summary}); detailed per-participant modality choices are provided in Appendix~\ref{app:strategies_results}. When adding new objects (Task~1), nearly all participants (27/30) combined text and visual input, indicating that introducing new content required both spatial specification and semantic clarification. Pattern addition (Task~3) relied mainly on visual input---18 participants used visuals exclusively and 12 combined them with text.  
Referencing attributes and orientation changes (Tasks~4--5) were dominated by visual input (typically arrows), with short text labels added for precision or reassurance. Moving objects (Task~7) showed a similar pattern, again favoring arrows for spatial changes. Only shape changes (Task~2) relied more on text, as participants preferred naming familiar objects (e.g., \textit{``double-decker bus''}) over drawing them. Overall, modality choice was strongly shaped by task type and object familiarity.

\paragraph{Workflow patterns.}  
Across tasks, most participants followed a \textit{visual-first workflow}: quick sketches or arrows for spatial intent, then text for clarification or detail. This hybrid coordination balanced speed and precision, enabling greater expressiveness, confidence, and control. While some participants noted friction when switching between stylus and keyboard (\textit{``Switching tools costs time and changes your mental model.''}, P4), others found the setup consistent with their everyday workflows (P16, P28).

\section{Discussion}
\label{sec:discussion}
This study examined how visual, textual, and multimodal input modalities influence GenAI image refinement. Beyond task-level results, this section interprets the broader implications of these findings for GenAI refinement. We discuss how visual--text integration balances expressiveness and efficiency, how multimodal workflows support iterative and creative refinement, and how coordination between inputs can minimize switching friction. We also consider differences across user expertise and contexts before concluding with a discussion of study limitations and directions for future work.

\subsection{Balancing Expressiveness and Efficiency Through Visual--Text Integration}
Visual input improved expressiveness by reducing ambiguity in describing spatial changes and object relationships, while lowering the cognitive effort required to specify details through text. Extending prior work on sketch-based and multimodal interfaces~\cite{lin_sketchflex_2025, shi_brickify_2025, peng_fusain_2025, weber_drawing--steps_2025, lin_inkspire_2025}, our findings show that scribbles and annotations convey layered spatial meaning rather than serving only as visual references. Participants used quick arrows, circles, and rough sketches to externalize intent that would otherwise be difficult to verbalize. Two sketching modes emerged: precise sketches for control and loose sketches for exploration, reflecting the dual role of ambiguity in design~\cite{gross_ambiguous_1996}. Future systems could include controls that adjust how strongly visual prompts influence model interpretation, supporting both concrete manipulation (e.g., repositioning or resizing) and exploratory edits (e.g., tone or style variations). 

Visual prompting not only clarified intent but also accelerated the refinement process. Quick visual expressions allowed designers to ``draw what they mean,'' avoiding the trial--and--error phrasing often required by text prompts~\cite{tholander_design_2023, park_we_2024}. Participants frequently combined visual input with short text to add semantic or stylistic detail without increasing task completion time. Quantitative results confirmed that the combined modality achieved efficiency comparable to visual input alone, suggesting that efficiency arises from fluid transitions between modalities rather than reliance on a single one. Importantly, these results show that the value of multimodal prompting lies not simply in providing more input, but in enabling task-appropriate coordination between modalities. While the combined modality received the highest overall ratings, it did not uniformly outperform single-modality input. For spatial refinement tasks, visual input alone was often sufficient and more direct, whereas text was most effective for semantic and global changes. This differentiation highlights that effective multimodal refinement depends on how modalities are used in relation to task demands, rather than on their simultaneous presence. Overall, these findings suggest that the observed patterns are influenced less by the particular GenAI models used and more by how different input modalities support the expression of refinement intent, reflecting interaction-level dynamics that are likely to persist across GenAI image systems with comparable capabilities.

\subsection{Iterative Refinement and Creative Flow} 
Multimodal input supported iterative refinement by balancing precise control with creative exploration. Visual input enabled rapid, direct manipulation without translating ideas into text, helping maintain creative flow. Participants often adopted a visual-first, text-second workflow: beginning with quick scribbles to externalize spatial intent and then layering text to clarify or extend meaning. This approach allowed stepwise iteration while retaining context between changes. This pattern indicates that designers actively choose and sequence modalities to manage ambiguity and maintain control during refinement, rather than defaulting to multimodal input simply because it is available. Such behavior aligns with prior accounts of sketch-based iteration, where rough sketches invite reinterpretation while detailed ones constrain outcomes~\cite{tang_comparing_2011, gross_ambiguous_1996, gero_exploration_2006, gaver_ambiguity_2003}.
These findings suggest that multimodal prompting reframes refinement as an interactive, evolving process rather than a linear sequence of corrections. Fluid transitions between modalities allowed participants' intent to develop in real time through reflection and reinterpretation. Designers switched modalities based on task type and desired control level. Interfaces that preserve this flow should keep interaction on the canvas, provide immediate visual feedback, and support small, reversible steps that encourage experimentation. Quantitative results reinforce this pattern, showing that the combined modality supported iterative refinement more effectively than either text or visual input alone. Multimodal workflows thus sustain creative momentum and enable continuous, context-aware refinement without disrupting rhythm.

\subsection{Multimodal Coordination and Minimizing Switching Friction}
When both input types were available, participants adopted two main strategies. In complementary use, scribbles conveyed spatial composition while text specified semantic or stylistic details. In redundant use, participants repeated information across modalities to confirm system interpretation.
Such redundancy could be interpreted as inefficiency or as a sign of insufficient system feedback and transparency. In our study, participants often responded to this uncertainty by deliberately repeating information across modalities as a way to gain confidence in how their input would be interpreted. In this sense, redundancy functioned as an adaptive strategy while also highlighting opportunities to improve system feedback and transparency. These strategies unfolded within a hybrid physical device workflow. Participants frequently alternated between drawing with a pen, typing on a keyboard, and selecting regions, and several reported friction when switching between input devices, consistent with costs observed in prior work on hybrid interfaces~\cite{foley_switching_2022}. Although the combined modality did not increase task completion time, moments of hesitation or re-confirmation emerged during these transitions, shaping how participants coordinated modalities and when they chose to reinforce intent through redundancy. 
These observations highlight the need for environments that allow handwriting, drawing, and typing within a single continuous workspace, provide clear and timely feedback about input interpretation, and offer lightweight on-canvas tools that reduce coordination costs while treating redundancy as a valid confirmation strategy. By making interpretation more transparent, future systems could minimize switching friction and reduce the need for redundant specification while preserving users' sense of control.

\subsection{Adapting to Expertise and Context}
Input preferences varied by task type, user background, and device familiarity. Drawing-oriented professionals and stylus users generally favored visual input, whereas participants more accustomed to keyboard--mouse workflows or text-based GenAI tools preferred typing. Some participants noted that visual input could also benefit non-design contexts such as casual photo editing, marketing, or communication tasks~\cite{samsung_galaxy_ai_use_2024}. 
These findings suggest that visual prompting holds potential beyond professional design, extending to broader creative and everyday applications. To accommodate such diversity, GenAI systems should adapt to users' device ecologies by providing optimized workflows for pen tablets, touchscreens, and keyboard--mouse setups, and by allowing users to pin frequently used tools near the canvas. Multimodal interfaces should further account for different expertise levels and usage contexts, enabling accessible and flexible GenAI-assisted refinement across professional and casual use.

\subsection{Limitations and Future Work}
This study has several limitations. 
Because the study was conducted in a controlled laboratory environment with predefined tasks and limited iteration cycles, it may not fully capture real-world GenAI refinement practices. While the open-ended phase supported more exploratory and self-directed use of multimodal input, practical constraints such as session duration and model latency limited the extent of iterative refinement observed. The system also provided a limited visual lexicon for user reference, which may have influenced how participants interpreted and used visual prompts.

While participants represented multiple design domains, about one-third came from UI/UX backgrounds and varied in their familiarity with stylus input. Designers less accustomed to stylus-based pen tablet interaction may have experienced a learning curve that affected performance and preferences. Input modalities in the prototype were coupled with physical input devices: text-only prompting relied on mouse-based inpainting, while visual and combined prompting used pen input. This coupling reflects how current GenAI tools are typically encountered in practice, but it may have influenced performance or user experience, particularly for spatial refinements. Accordingly, the results should be interpreted as reflecting realistic interaction contexts rather than device-controlled comparisons, and systematically disentangling modality and device effects remains an important direction for future work.

The prototype also imposed technical constraints. Image quality occasionally degraded across repeated \textit{FLUX.1} iterations, and \textit{GPT-Image-1} incurred higher latency for complex tasks. Outputs were non-deterministic and sometimes produced minor inconsistencies across runs. Although participants were instructed to focus on interaction experience and modality expressiveness rather than output quality or speed, these factors may still have influenced perceptions of control and satisfaction. As model performance improves, faster and more stable responses may reveal further benefits of multimodal prompting. The prototype was instantiated using specific GenAI image models and a predefined set of seven refinement tasks. These tasks were derived from a formative study and represent common refinement actions in design practice, such as adding, moving, resizing, and modifying visual attributes. Accordingly, the findings should not be interpreted as benchmarks of model performance or as exhaustive coverage of refinement tasks, but rather as illustrations of how different input modalities shape refinement behavior within the studied setup. Moreover, the study did not include an objective measure of task success or output quality. While we assessed efficiency and subjective perceptions such as workload and perceived expressiveness, the quality of the final images was not evaluated through expert ratings or perceptual evaluation methods. Given the non-deterministic nature of GenAI outputs and the use of multiple models, introducing a single objective quality metric was beyond the scope of this work. Future research could apply expert-based or perceptual evaluation methods to directly assess output quality and examine potential speed--accuracy trade-offs across input modalities. 

Future work should involve a broader participant pool with more experienced pen users and conduct longitudinal or in-the-wild studies to observe multimodal refinement in natural GenAI use contexts. Such studies could also reveal richer visual languages of scribbles and annotations that communicate intent to GenAI systems. Extending beyond individual refinement workflows to collaborative scenarios could further show how scribbles and annotations support co-creation and shared intent~\cite{tao_designweaver_2025}. Future research could also expand beyond short, controlled refinement tasks to explore how multimodal prompting supports more dynamic or multi-stage creation (e.g., sequence edits or storytelling), improving ecological validity and relevance to real-world GenAI workflows. Finally, future work could investigate interface mechanisms that support smoother modality transitions, building on our findings about switching friction to explore how unified input systems can better sustain creative flow.

\section{Conclusion}
This work investigated how multimodal prompting---combining text, pen-based scribbles, and annotations---can enhance refinement in GenAI image tools. While existing systems primarily support ideation, we focused on iterative adjustments within GenAI workflows, where designers refine composition, orientation, visual details, and reference objects in images. Building on insights from a formative study with professional designers ($N = 7$), we developed a prototype and conducted a within-subjects study with 30 designers and design students to compare text-only, visual-only, and combined input modalities across both closed- and open-ended refinement tasks.
Our findings show that visual prompting offers clear advantages for expressing spatial and relational intent, improving clarity, speed, and perceived control in refinement. Text input, in contrast, remained effective for semantic and stylistic modifications that benefit from linguistic precision and conceptual flexibility. The combined modality achieved the highest overall ratings, enabling designers to balance spatial expressiveness with semantic detail through complementary and iterative use of both modalities. Participants frequently integrated visual and textual cues---either sequentially or redundantly---to reduce ambiguity and build confidence in their iterative workflows.
These results highlight how multimodal prompting can bridge the gap between text-centric GenAI interfaces and the visually grounded nature of design practice. By enabling designers to communicate refinement intent more directly and intuitively, such tools can evolve from rapid ideation aids into integrated supports for diverse stages of creative work. Future research could explore adaptive, task-aware interfaces that dynamically support modality switching with minimal friction, as well as longitudinal or in-the-wild studies to examine how multimodal prompting fosters creativity and supports diverse real-world GenAI workflows.

\begin{acks}
We sincerely thank all participants in our studies as well as the anonymous reviewers for their time and valuable feedback. We also thank Leo Leinmüller and Christian Schick for their support during the prototyping phase. For this manuscript, we used generative AI tools to improve readability (e.g., spelling and rephrasing). All substantive content, analysis, and conclusions were developed by the authors, who retain full responsibility for the work.
\end{acks}

\bibliographystyle{ACM-Reference-Format}
\bibliography{reference_cr}

@inproceedings{adamkiewicz_promptmap_2025,
	title        = {{PromptMap}: {An} {Alternative} {Interaction} {Style} for {AI}-{Based} {Image} {Generation}},
	shorttitle   = {{PromptMap}},
	author       = {Adamkiewicz, Krzysztof and Woźniak, Paweł Wojciech and Dominiak, Julia and Romanowski, Andrzej and Karolus, Jakob and Frolov, Stanislav},
	year         = 2025,
	month        = mar,
	booktitle    = {Proceedings of the 30th {International} {Conference} on {Intelligent} {User} {Interfaces}},
	publisher    = {ACM},
	address      = {New York, NY, USA},
	series       = {{IUI} '25},
	pages        = {1162--1176},
	doi          = {10.1145/3708359.3712150},
	isbn         = {979-8-4007-1306-4},
	url          = {https://dl.acm.org/doi/10.1145/3708359.3712150},
	urldate      = {2025-04-04},
	language     = {en-US}
}

@misc{adobe_adobe_2025,
	title        = {Adobe Firefly -- Generative {AI} for creatives},
	author       = {{Adobe}},
	year         = 2025,
	url          = {https://www.adobe.com/products/firefly.html},
	urldate      = {2026-01-15},
	language     = {en-US}
}

@misc{samsung_galaxy_ai_use_2024,
	title        = {Use {AI} editing tools in {Gallery} on your {Galaxy} phone or tablet},
	author       = {{Samsung}},
	year         = 2024,
	url          = {https://www.samsung.com/us/support/answer/ANS10000934/},
	urldate      = {2026-01-15},
	language     = {en-US}
}

@inproceedings{plimmer_computer-aided_2002,
	title        = {Computer-aided sketching to capture preliminary design},
	author       = {Plimmer, Beryl and Apperley, Mark},
	year         = 2002,
	booktitle    = {Proceedings of the Third Australasian User Interfaces Conference (AUIC2002)},
	publisher    = {Australian Computer Society, Inc.},
	address      = {Melbourne, Australia},
	series       = {Conferences in Research and Practice in Information Technology},
	volume       = 7,
	url          = {https://hdl.handle.net/10289/4721},
	urldate      = {2026-01-15},
	editor       = {Grundy, John and Calder, Paul},
	language     = {en}
}

@inproceedings{brade_promptify_2023,
	title        = {Promptify: {Text}-to-{Image} {Generation} through {Interactive} {Prompt} {Exploration} with {Large} {Language} {Models}},
	shorttitle   = {Promptify},
	author       = {Brade, Stephen and Wang, Bryan and Sousa, Mauricio and Oore, Sageev and Grossman, Tovi},
	year         = 2023,
	month        = oct,
	booktitle    = {Proceedings of the 36th {Annual} {ACM} {Symposium} on {User} {Interface} {Software} and {Technology}},
	publisher    = {ACM},
	address      = {New York, NY, USA},
	series       = {{UIST} '23},
	pages        = {1--14},
	doi          = {10.1145/3586183.3606725},
	isbn         = {979-8-4007-0132-0},
	url          = {https://dl.acm.org/doi/10.1145/3586183.3606725},
	urldate      = {2024-08-21},
	language     = {en-US}
}

@article{braun_one_2021,
	title        = {One size fits all? {What} counts as quality practice in (reflexive) thematic analysis?},
	shorttitle   = {One size fits all?},
	author       = {Braun, Virginia and Clarke, Victoria},
	year         = 2021,
	month        = jul,
	journal      = {Qualitative Research in Psychology},
	volume       = 18,
	number       = 3,
	pages        = {328--352},
	doi          = {10.1080/14780887.2020.1769238},
	issn         = {1478-0887, 1478-0895},
	url          = {https://www.tandfonline.com/doi/full/10.1080/14780887.2020.1769238},
	urldate      = {2024-07-31},
	language     = {en}
}

@inproceedings{chen_autospark_2024,
	title        = {{AutoSpark}: {Supporting} {Automobile} {Appearance} {Design} {Ideation} with {Kansei} {Engineering} and {Generative} {AI}},
	shorttitle   = {{AutoSpark}},
	author       = {Chen, Liuqing and Jing, Qianzhi and Tsang, Yixin and Wang, Qianyi and Liu, Ruocong and Xia, Duowei and Zhou, Yunzhan and Sun, Lingyun},
	year         = 2024,
	month        = oct,
	booktitle    = {Proceedings of the 37th {Annual} {ACM} {Symposium} on {User} {Interface} {Software} and {Technology}},
	publisher    = {ACM},
	address      = {Pittsburgh PA USA},
	pages        = {1--19},
	doi          = {10.1145/3654777.3676337},
	isbn         = {979-8-4007-0628-8},
	url          = {https://dl.acm.org/doi/10.1145/3654777.3676337},
	urldate      = {2024-12-23},
	language     = {en}
}

@article{chen_how_2025,
	title        = {How generative {AI} supports human in conceptual design},
	author       = {Chen, Liuqing and Song, Yaxuan and Guo, Jia and Sun, Lingyun and Childs, Peter and Yin, Yuan},
	year         = 2025,
	journal      = {Design Science},
	volume       = 11,
	pages        = {e9},
	doi          = {10.1017/dsj.2025.2},
	issn         = {2053-4701},
	url          = {https://www.cambridge.org/core/product/identifier/S2053470125000022/type/journal_article},
	urldate      = {2025-09-02},
	language     = {en}
}

@inproceedings{choi_creativeconnect_2024,
	title        = {{CreativeConnect}: {Supporting} {Reference} {Recombination} for {Graphic} {Design} {Ideation} with {Generative} {AI}},
	shorttitle   = {{CreativeConnect}},
	author       = {Choi, DaEun and Hong, Sumin and Park, Jeongeon and Chung, John Joon Young and Kim, Juho},
	year         = 2024,
	month        = may,
	booktitle    = {Proceedings of the {CHI} {Conference} on {Human} {Factors} in {Computing} {Systems}},
	publisher    = {ACM},
	address      = {Honolulu HI USA},
	pages        = {1--25},
	doi          = {10.1145/3613904.3642794},
	isbn         = {979-8-4007-0330-0},
	url          = {https://dl.acm.org/doi/10.1145/3613904.3642794},
	urldate      = {2024-07-10},
	language     = {en}
}

@inproceedings{choi_expandora_2025,
	title        = {Expandora: {Broadening} {Design} {Exploration} with {Text}-to-{Image} {Model}},
	shorttitle   = {Expandora},
	author       = {Choi, DaEun and Son, Kihoon and Jung, HyunJoon and Kim, Juho},
	year         = 2025,
	month        = apr,
	booktitle    = {Proceedings of the {Extended} {Abstracts} of the {CHI} {Conference} on {Human} {Factors} in {Computing} {Systems}},
	publisher    = {ACM},
	address      = {New York, NY, USA},
	series       = {{CHI} {EA} '25},
	pages        = {1--10},
	doi          = {10.1145/3706599.3720189},
	isbn         = {979-8-4007-1395-8},
	url          = {https://dl.acm.org/doi/10.1145/3706599.3720189},
	urldate      = {2025-05-28}
}

@inproceedings{chung_promptpaint_2023,
	title        = {{PromptPaint}: {Steering} {Text}-to-{Image} {Generation} {Through} {Paint} {Medium}-like {Interactions}},
	shorttitle   = {{PromptPaint}},
	author       = {Chung, John Joon Young and Adar, Eytan},
	year         = 2023,
	month        = oct,
	booktitle    = {Proceedings of the 36th {Annual} {ACM} {Symposium} on {User} {Interface} {Software} and {Technology}},
	publisher    = {ACM},
	address      = {San Francisco CA USA},
	pages        = {1--17},
	doi          = {10.1145/3586183.3606777},
	isbn         = {979-8-4007-0132-0},
	url          = {https://dl.acm.org/doi/10.1145/3586183.3606777},
	urldate      = {2024-04-07},
	language     = {en}
}

@misc{dalle_3_dalle_2025,
	title        = {{DALL-E} 3},
	author       = {{OpenAI}},
	year         = 2025,
	url          = {https://openai.com/index/dall-e-3/},
	urldate      = {2026-01-15},
	language     = {en-US}
}

@inproceedings{dang_worldsmith_2023,
	title        = {{WorldSmith}: {Iterative} and {Expressive} {Prompting} for {World} {Building} with a {Generative} {AI}},
	shorttitle   = {{WorldSmith}},
	author       = {Dang, Hai and Brudy, Frederik and Fitzmaurice, George and Anderson, Fraser},
	year         = 2023,
	month        = oct,
	booktitle    = {Proceedings of the 36th {Annual} {ACM} {Symposium} on {User} {Interface} {Software} and {Technology}},
	publisher    = {ACM},
	address      = {New York, NY, USA},
	series       = {{UIST} '23},
	pages        = {1--17},
	doi          = {10.1145/3586183.3606772},
	isbn         = {979-8-4007-0132-0},
	url          = {https://dl.acm.org/doi/10.1145/3586183.3606772},
	urldate      = {2024-09-29},
	language     = {en-US}
}

@inproceedings{feng_how_2023,
	title        = {How {Do} {UX} {Practitioners} {Communicate} {AI} as a {Design} {Material}? {Artifacts}, {Conceptions}, and {Propositions}},
	shorttitle   = {How {Do} {UX} {Practitioners} {Communicate} {AI} as a {Design} {Material}?},
	author       = {Feng, K. J. Kevin and Coppock, Maxwell James and McDonald, David W.},
	year         = 2023,
	month        = jul,
	booktitle    = {Proceedings of the 2023 {ACM} {Designing} {Interactive} {Systems} {Conference}},
	publisher    = {ACM},
	address      = {Pittsburgh PA USA},
	pages        = {2263--2280},
	doi          = {10.1145/3563657.3596101},
	isbn         = {978-1-4503-9893-0},
	url          = {https://dl.acm.org/doi/10.1145/3563657.3596101},
	urldate      = {2024-01-28},
	language     = {en}
}

@inproceedings{foley_switching_2022,
	title        = {Switching {Between} {Standard} {Pointing} {Methods} with {Current} and {Emerging} {Computer} {Form} {Factors}},
	author       = {Foley, Margaret Jean and Roy, Quentin and Huang, Da-Yuan and Li, Wei and Vogel, Daniel},
	year         = 2022,
	month        = apr,
	booktitle    = {Proceedings of the 2022 {CHI} {Conference} on {Human} {Factors} in {Computing} {Systems}},
	publisher    = {ACM},
	address      = {New York, NY, USA},
	series       = {{CHI} '22},
	pages        = {1--14},
	doi          = {10.1145/3491102.3517433},
	isbn         = {978-1-4503-9157-3},
	url          = {https://dl.acm.org/doi/10.1145/3491102.3517433},
	urldate      = {2025-10-06}
}

@inproceedings{gmeiner_exploring_2023,
	title        = {Exploring {Challenges} and {Opportunities} to {Support} {Designers} in {Learning} to {Co}-create with {AI}-based {Manufacturing} {Design} {Tools}},
	author       = {Gmeiner, Frederic and Yang, Humphrey and Yao, Lining and Holstein, Kenneth and Martelaro, Nikolas},
	year         = 2023,
	month        = apr,
	booktitle    = {Proceedings of the 2023 {CHI} {Conference} on {Human} {Factors} in {Computing} {Systems}},
	publisher    = {ACM},
	address      = {Hamburg Germany},
	pages        = {1--20},
	doi          = {10.1145/3544548.3580999},
	isbn         = {978-1-4503-9421-5},
	url          = {https://dl.acm.org/doi/10.1145/3544548.3580999},
	urldate      = {2024-02-06},
	language     = {en}
}

@article{goldschmidt_backtalk_2003,
	title        = {The {Backtalk} of {Self}-{Generated} {Sketches}},
	author       = {Goldschmidt, Gabriela},
	year         = 2003,
	journal      = {Design Issues},
	publisher    = {MIT Press},
	volume       = 19,
	number       = 1,
	pages        = {72--88},
	issn         = {0747-9360},
	url          = {https://www.jstor.org/stable/1512057},
	urldate      = {2025-09-20}
}

@inproceedings{gross_ambiguous_1996,
	title        = {Ambiguous intentions: a paper-like interface for creative design},
	shorttitle   = {Ambiguous intentions},
	author       = {Gross, Mark D. and Do, Ellen Yi-Luen},
	year         = 1996,
	booktitle    = {Proceedings of the 9th annual {ACM} symposium on {User} interface software and technology  - {UIST} '96},
	publisher    = {ACM},
	address      = {Seattle, Washington, United States},
	pages        = {183--192},
	doi          = {10.1145/237091.237119},
	isbn         = {978-0-89791-798-8},
	url          = {https://dl.acm.org/doi/10.1145/237091.237119},
	urldate      = {2024-04-29},
	language     = {en}
}

@inproceedings{hinckley_pen_2010,
	title        = {Pen + touch = new tools},
	author       = {Hinckley, Ken and Yatani, Koji and Pahud, Michel and Coddington, Nicole and Rodenhouse, Jenny and Wilson, Andy and Benko, Hrvoje and Buxton, Bill},
	year         = 2010,
	month        = oct,
	booktitle    = {Proceedings of the 23rd annual {ACM} symposium on {User} interface software and technology},
	publisher    = {ACM},
	address      = {New York New York USA},
	pages        = {27--36},
	doi          = {10.1145/1866029.1866036},
	isbn         = {978-1-4503-0271-5},
	url          = {https://dl.acm.org/doi/10.1145/1866029.1866036},
	urldate      = {2025-09-30},
	language     = {en}
}

@inproceedings{huang_plantography_2024,
	title        = {{PlantoGraphy}: {Incorporating} {Iterative} {Design} {Process} into {Generative} {Artificial} {Intelligence} for {Landscape} {Rendering}},
	shorttitle   = {{PlantoGraphy}},
	author       = {Huang, Rong and Lin, Haichuan and Chen, Chuanzhang and Zhang, Kang and Zeng, Wei},
	year         = 2024,
	month        = may,
	booktitle    = {Proceedings of the {CHI} {Conference} on {Human} {Factors} in {Computing} {Systems}},
	publisher    = {ACM},
	address      = {Honolulu HI USA},
	pages        = {1--19},
	doi          = {10.1145/3613904.3642824},
	isbn         = {979-8-4007-0330-0},
	url          = {https://dl.acm.org/doi/10.1145/3613904.3642824},
	urldate      = {2024-05-15},
	language     = {en}
}

@inproceedings{inie_designing_2023,
	title        = {Designing {Participatory} {AI}: {Creative} {Professionals}’ {Worries} and {Expectations} about {Generative} {AI}},
	shorttitle   = {Designing {Participatory} {AI}},
	author       = {Inie, Nanna and Falk, Jeanette and Tanimoto, Steve},
	year         = 2023,
	month        = apr,
	booktitle    = {Extended {Abstracts} of the 2023 {CHI} {Conference} on {Human} {Factors} in {Computing} {Systems}},
	publisher    = {ACM},
	address      = {Hamburg Germany},
	pages        = {1--8},
	doi          = {10.1145/3544549.3585657},
	isbn         = {978-1-4503-9422-2},
	url          = {https://dl.acm.org/doi/10.1145/3544549.3585657},
	urldate      = {2024-04-24},
	language     = {en}
}

@misc{dalle_inpainting_editing_2024,
	title        = {Editing your images with {DALL-E} | {OpenAI} Help Center},
	author       = {{OpenAI}},
	year         = 2024,
	url          = {https://help.openai.com/en/articles/9055440-editing-your-images-with-dall-e},
	urldate      = {2026-01-15},
	language     = {en-US}
}

@article{jonson_design_2005,
	title        = {Design ideation: the conceptual sketch in the digital age},
	shorttitle   = {Design ideation},
	author       = {Jonson, Ben},
	year         = 2005,
	month        = nov,
	journal      = {Design Studies},
	volume       = 26,
	number       = 6,
	pages        = {613--624},
	doi          = {10.1016/j.destud.2005.03.001},
	issn         = {0142-694X},
	url          = {https://www.sciencedirect.com/science/article/pii/S0142694X05000189},
	urldate      = {2024-12-30}
}

@inproceedings{kim_shoegenai_2025,
	title        = {{ShoeGenAI}: {A} {Creativity} {Support} {Tool} {Bridging} {Design} {Intention} and {Feasibility} in {Shoe} {Design}},
	shorttitle   = {{ShoeGenAI}},
	author       = {Kim, Hui-Jun and Kim, Jeongho and Jeong, Sohyun and Lee, Minbong and Choo, Jaegul and Kim, Sung-Hee},
	year         = 2025,
	month        = sep,
	booktitle    = {Proceedings of the 38th {Annual} {ACM} {Symposium} on {User} {Interface} {Software} and {Technology}},
	publisher    = {ACM},
	address      = {New York, NY, USA},
	series       = {{UIST} '25},
	pages        = {1--18},
	doi          = {10.1145/3746059.3747691},
	isbn         = {979-8-4007-2037-6},
	url          = {https://dl.acm.org/doi/10.1145/3746059.3747691},
	urldate      = {2025-09-29}
}

@inproceedings{lee_impact_2024,
	title        = {The {Impact} of {Sketch}-guided vs. {Prompt}-guided {3D} {Generative} {AIs} on the {Design} {Exploration} {Process}},
	author       = {Lee, Seung Won and Jo, Tae Hee and Jin, Semin and Choi, Jiin and Yun, Kyungwon and Bromberg, Sergio and Ban, Seonghoon and Hyun, Kyung Hoon},
	year         = 2024,
	month        = may,
	booktitle    = {Proceedings of the {CHI} {Conference} on {Human} {Factors} in {Computing} {Systems}},
	publisher    = {ACM},
	address      = {Honolulu HI USA},
	pages        = {1--18},
	doi          = {10.1145/3613904.3642218},
	isbn         = {979-8-4007-0330-0},
	url          = {https://dl.acm.org/doi/10.1145/3613904.3642218},
	urldate      = {2024-05-15},
	language     = {en}
}

@article{li_realtimegen_2025,
	title        = {{RealtimeGen}: {An} {Intervenable} {AI} {Image} {Generation} {System} for {Commercial} {Digital} {Art} {Asset} {Creators}},
	shorttitle   = {{RealtimeGen}},
	author       = {Li, Zejian and Zhang, Ying and Zhou, Shengzhe and Liu, Qi and Zhang, Jiesi and Xu, Haoran and Chen, Shuyao and Chen, Xiaoyu and Sun, Lingyun},
	year         = 2025,
	month        = jun,
	journal      = {International Journal of Human–Computer Interaction},
	publisher    = {Taylor \& Francis},
	volume       = 41,
	number       = 11,
	pages        = {6613--6636},
	doi          = {10.1080/10447318.2024.2382508},
	issn         = {1044-7318},
	url          = {https://doi.org/10.1080/10447318.2024.2382508},
	urldate      = {2025-08-30},
	note         = {\_eprint: https://doi.org/10.1080/10447318.2024.2382508}
}

@inproceedings{liang_storydiffusion_2025,
	title        = {{StoryDiffusion}: {How} to {Support} {UX} {Storyboarding} {With} {Generative}-{AI}},
	shorttitle   = {{StoryDiffusion}},
	author       = {Liang, Zhaohui and Zhang, Xiaoyu and Ma, Kevin and Liu, Zhao and Ren, Xipei and Goucher-Lambert, Kosa and Liu, Can},
	year         = 2025,
	month        = oct,
	booktitle    = {Proceedings of the 27th {International} {Conference} on {Multimodal} {Interaction}},
	publisher    = {ACM},
	address      = {New York, NY, USA},
	series       = {{ICMI} '25},
	pages        = {135--144},
	doi          = {10.1145/3716553.3750793},
	isbn         = {979-8-4007-1499-3},
	url          = {https://dl.acm.org/doi/10.1145/3716553.3750793},
	urldate      = {2026-01-20}
}

@inproceedings{lin_inkspire_2025,
	title        = {Inkspire: {Supporting} {Design} {Exploration} with {Generative} {AI} through {Analogical} {Sketching}},
	shorttitle   = {Inkspire},
	author       = {Lin, David Chuan-En and Kang, Hyeonsu B. and Martelaro, Nikolas and Kittur, Aniket and Chen, Yan-Ying and Hong, Matthew K.},
	year         = 2025,
	month        = apr,
	booktitle    = {Proceedings of the 2025 {CHI} {Conference} on {Human} {Factors} in {Computing} {Systems}},
	publisher    = {ACM},
	address      = {New York, NY, USA},
	series       = {{CHI} '25},
	pages        = {1--18},
	doi          = {10.1145/3706598.3713397},
	isbn         = {979-8-4007-1394-1},
	url          = {https://dl.acm.org/doi/10.1145/3706598.3713397},
	urldate      = {2025-06-14},
	language     = {en-US}
}

@inproceedings{lin_sketchflex_2025,
	title        = {{SketchFlex}: {Facilitating} {Spatial}-{Semantic} {Coherence} in {Text}-to-{Image} {Generation} with {Region}-{Based} {Sketches}},
	shorttitle   = {{SketchFlex}},
	author       = {Lin, Haichuan and Ye, Yilin and Xia, Jiazhi and Zeng, Wei},
	year         = 2025,
	month        = apr,
	booktitle    = {Proceedings of the 2025 {CHI} {Conference} on {Human} {Factors} in {Computing} {Systems}},
	publisher    = {ACM},
	address      = {New York, NY, USA},
	series       = {{CHI} '25},
	pages        = {1--19},
	doi          = {10.1145/3706598.3713801},
	isbn         = {979-8-4007-1394-1},
	url          = {https://dl.acm.org/doi/10.1145/3706598.3713801},
	urldate      = {2025-08-30},
	language     = {en-US}
}

@inproceedings{liu_opal_2022,
	title        = {Opal: {Multimodal} {Image} {Generation} for {News} {Illustration}},
	shorttitle   = {Opal},
	author       = {Liu, Vivian and Qiao, Han and Chilton, Lydia},
	year         = 2022,
	month        = oct,
	booktitle    = {Proceedings of the 35th {Annual} {ACM} {Symposium} on {User} {Interface} {Software} and {Technology}},
	publisher    = {ACM},
	address      = {New York, NY, USA},
	series       = {{UIST} '22},
	pages        = {1--17},
	doi          = {10.1145/3526113.3545621},
	isbn         = {978-1-4503-9320-1},
	url          = {https://dl.acm.org/doi/10.1145/3526113.3545621},
	urldate      = {2024-08-23}
}

@inproceedings{liu_3dall-e_2023,
	title        = {{3DALL}-{E}: {Integrating} {Text}-to-{Image} {AI} in {3D} {Design} {Workflows}},
	shorttitle   = {{3DALL}-{E}},
	author       = {Liu, Vivian and Vermeulen, Jo and Fitzmaurice, George and Matejka, Justin},
	year         = 2023,
	month        = jul,
	booktitle    = {Proceedings of the 2023 {ACM} {Designing} {Interactive} {Systems} {Conference}},
	publisher    = {ACM},
	address      = {Pittsburgh PA USA},
	pages        = {1955--1977},
	doi          = {10.1145/3563657.3596098},
	isbn         = {978-1-4503-9893-0},
	url          = {https://dl.acm.org/doi/10.1145/3563657.3596098},
	urldate      = {2024-04-29},
	language     = {en}
}

@misc{lllyasviel_lllyasvielcontrolnet_2025,
	title        = {{ControlNet}},
	author       = {{lllyasviel}},
	year         = 2025,
	url          = {https://github.com/lllyasviel/ControlNet},
	urldate      = {2026-01-15},
	note         = {GitHub repository (Apache-2.0 license)}
}

@inproceedings{michelessa_varifai_2025,
	title        = {Varif.ai to {Vary} and {Verify} {User}-{Driven} {Diversity} in {Scalable} {Image} {Generation}},
	author       = {Michelessa, Mario and Ng, Jamie and Hurter, Christophe and Lim, Brian Y.},
	year         = 2025,
	month        = jul,
	booktitle    = {Proceedings of the 2025 {ACM} {Designing} {Interactive} {Systems} {Conference}},
	publisher    = {ACM},
	address      = {New York, NY, USA},
	series       = {{DIS} '25},
	pages        = {1867--1885},
	doi          = {10.1145/3715336.3735847},
	isbn         = {979-8-4007-1485-6},
	url          = {https://dl.acm.org/doi/10.1145/3715336.3735847},
	urldate      = {2025-07-29},
	language     = {en-US}
}

@misc{midjourney_midjourney_2025,
	title        = {Midjourney},
	author       = {{Midjourney}},
	year         = 2025,
	url          = {https://www.midjourney.com/website},
	urldate      = {2026-01-15}
}

@misc{park_designing_2024,
	title        = {Designing for {Visual} {Thinkers}: {Overcoming} {Text}-{Centric} {Limitations} in {GenAI} {Tools}},
	author       = {Park, Hyerim and Eiband, Malin},
	year         = 2024,
	month        = oct,
	url          = {https://doi.org/10.5281/zenodo.14186390},
	urldate      = {2026-01-15},
	note         = {Workshop paper, Zenodo}
}

@inproceedings{park_we_2024,
	title        = {"{We} {Are} {Visual} {Thinkers}, {Not} {Verbal} {Thinkers}!": {A} {Thematic} {Analysis} of {How} {Professional} {Designers} {Use} {Generative} {AI} {Image} {Generation} {Tools}},
	shorttitle   = {"{We} {Are} {Visual} {Thinkers}, {Not} {Verbal} {Thinkers}!"},
	author       = {Park, Hyerim and Eirich, Joscha and Luckow, Andre and Sedlmair, Michael},
	year         = 2024,
	month        = oct,
	booktitle    = {Nordic {Conference} on {Human}-{Computer} {Interaction}},
	publisher    = {ACM},
	address      = {Uppsala Sweden},
	pages        = {1--14},
	doi          = {10.1145/3679318.3685370},
	isbn         = {979-8-4007-0966-1},
	url          = {https://dl.acm.org/doi/10.1145/3679318.3685370},
	urldate      = {2024-10-23},
	language     = {en}
}

@inproceedings{park_design_2025,
	title        = {Design and Evaluation of a Generative {AI}-Driven {VR} Texturing Tool: A Design Science Approach},
	author       = {Park, Hyerim and Eirich, Joscha and Luckow, Andre and Sedlmair, Michael},
	year         = 2025,
	booktitle    = {ECIS 2025 Proceedings},
	publisher    = {Association for Information Systems},
	url          = {https://aisel.aisnet.org/ecis2025/hci/hci/2},
	urldate      = {2026-01-20}
}

@inproceedings{peng_designprompt_2024,
	title        = {{DesignPrompt}: {Using} {Multimodal} {Interaction} for {Design} {Exploration} with {Generative} {AI}},
	shorttitle   = {{DesignPrompt}},
	author       = {Peng, Xiaohan and Koch, Janin and Mackay, Wendy E.},
	year         = 2024,
	month        = jul,
	booktitle    = {Designing {Interactive} {Systems} {Conference}},
	publisher    = {ACM},
	address      = {IT University of Copenhagen Denmark},
	pages        = {804--818},
	doi          = {10.1145/3643834.3661588},
	isbn         = {979-8-4007-0583-0},
	url          = {https://dl.acm.org/doi/10.1145/3643834.3661588},
	urldate      = {2024-07-09},
	language     = {en}
}

@inproceedings{peng_fusain_2025,
	title        = {{FusAIn}: {Composing} {Generative} {AI} {Visual} {Prompts} {Using} {Pen}-based {Interaction}},
	shorttitle   = {{FusAIn}},
	author       = {Peng, Xiaohan and Koch, Janin and Mackay, Wendy E.},
	year         = 2025,
	month        = apr,
	booktitle    = {Proceedings of the 2025 {CHI} {Conference} on {Human} {Factors} in {Computing} {Systems}},
	publisher    = {ACM},
	address      = {New York, NY, USA},
	series       = {{CHI} '25},
	pages        = {1--20},
	doi          = {10.1145/3706598.3714027},
	isbn         = {979-8-4007-1394-1},
	url          = {https://dl.acm.org/doi/10.1145/3706598.3714027},
	urldate      = {2025-04-28},
	language     = {en-US}
}

@article{schon_designing_1992,
	title        = {Designing as reflective conversation with the materials of a design situation},
	author       = {Schön, D. A.},
	year         = 1992,
	month        = mar,
	journal      = {Knowledge-Based Systems},
	series       = {Artificial {Intelligence} in {Design} {Conference} 1991 {Special} {Issue}},
	volume       = 5,
	number       = 1,
	pages        = {3--14},
	doi          = {10.1016/0950-7051(92)90020-G},
	issn         = {0950-7051},
	url          = {https://www.sciencedirect.com/science/article/pii/095070519290020G},
	urldate      = {2025-09-30}
}

@inproceedings{shi_brickify_2025,
	title        = {Brickify: {Enabling} {Expressive} {Design} {Intent} {Specification} through {Direct} {Manipulation} on {Design} {Tokens}},
	shorttitle   = {Brickify},
	author       = {Shi, Xinyu and Wang, Yinghou and Rossi, Ryan and Zhao, Jian},
	year         = 2025,
	month        = apr,
	booktitle    = {Proceedings of the 2025 {CHI} {Conference} on {Human} {Factors} in {Computing} {Systems}},
	publisher    = {ACM},
	address      = {Yokohama Japan},
	pages        = {1--20},
	doi          = {10.1145/3706598.3714087},
	isbn         = {979-8-4007-1394-1},
	url          = {https://dl.acm.org/doi/10.1145/3706598.3714087},
	urldate      = {2025-05-02},
	language     = {en}
}

@article{shi_understanding_2023,
	title        = {Understanding {Design} {Collaboration} {Between} {Designers} and {Artificial} {Intelligence}: {A} {Systematic} {Literature} {Review}},
	shorttitle   = {Understanding {Design} {Collaboration} {Between} {Designers} and {Artificial} {Intelligence}},
	author       = {Shi, Yang and Gao, Tian and Jiao, Xiaohan and Cao, Nan},
	year         = 2023,
	month        = oct,
	journal      = {Proc. ACM Hum.-Comput. Interact.},
	volume       = 7,
	number       = {CSCW2},
	pages        = {368:1--368:35},
	doi          = {10.1145/3610217},
	url          = {https://dl.acm.org/doi/10.1145/3610217},
	urldate      = {2025-01-29},
	language     = {en-US}
}

@inproceedings{son_genquery_2024,
	title        = {{GenQuery}: {Supporting} {Expressive} {Visual} {Search} with {Generative} {Models}},
	shorttitle   = {{GenQuery}},
	author       = {Son, Kihoon and Choi, DaEun and Kim, Tae Soo and Kim, Young-Ho and Kim, Juho},
	year         = 2024,
	month        = may,
	booktitle    = {Proceedings of the {CHI} {Conference} on {Human} {Factors} in {Computing} {Systems}},
	publisher    = {ACM},
	address      = {New York, NY, USA},
	series       = {{CHI} '24},
	pages        = {1--19},
	doi          = {10.1145/3613904.3642847},
	isbn         = {979-8-4007-0330-0},
	url          = {https://dl.acm.org/doi/10.1145/3613904.3642847},
	urldate      = {2024-08-29},
	language     = {en-US}
}

@inproceedings{sun_creative_2025,
	title        = {Creative {Blends} of {Visual} {Concepts}},
	author       = {Sun, Zhida and Zhang, Zhenyao and Zhang, Yue and Lu, Min and Lischinski, Dani and Cohen-Or, Daniel and Huang, Hui},
	year         = 2025,
	month        = apr,
	booktitle    = {Proceedings of the 2025 {CHI} {Conference} on {Human} {Factors} in {Computing} {Systems}},
	publisher    = {ACM},
	address      = {New York, NY, USA},
	series       = {{CHI} '25},
	pages        = {1--17},
	doi          = {10.1145/3706598.3713683},
	isbn         = {979-8-4007-1394-1},
	url          = {https://doi.org/10.1145/3706598.3713683},
	urldate      = {2025-06-25},
	language     = {en-US}
}

@article{tang_comparing_2011,
	title        = {Comparing collaborative co-located and distributed design processes in digital and traditional sketching environments: {A} protocol study using the function–behaviour–structure coding scheme},
	shorttitle   = {Comparing collaborative co-located and distributed design processes in digital and traditional sketching environments},
	author       = {Tang, H. H. and Lee, Y. Y. and Gero, J. S.},
	year         = 2011,
	month        = jan,
	journal      = {Design Studies},
	volume       = 32,
	number       = 1,
	pages        = {1--29},
	doi          = {10.1016/j.destud.2010.06.004},
	issn         = {0142-694X},
	url          = {https://www.sciencedirect.com/science/article/pii/S0142694X10000475},
	urldate      = {2025-10-06}
}

@inproceedings{tang_exploring_2024,
	title        = {Exploring the {Impact} of {AI}-generated {Image} {Tools} on {Professional} and {Non}-professional {Users} in the {Art} and {Design} {Fields}},
	author       = {Tang, Yuying and Zhang, Ningning and Ciancia, Mariana and Wang, Zhigang},
	year         = 2024,
	month        = nov,
	booktitle    = {Companion {Publication} of the 2024 {Conference} on {Computer}-{Supported} {Cooperative} {Work} and {Social} {Computing}},
	publisher    = {ACM},
	address      = {New York, NY, USA},
	series       = {{CSCW} {Companion} '24},
	pages        = {451--458},
	doi          = {10.1145/3678884.3681890},
	isbn         = {979-8-4007-1114-5},
	url          = {https://dl.acm.org/doi/10.1145/3678884.3681890},
	urldate      = {2024-11-24}
}

@inproceedings{tao_designweaver_2025,
	title        = {{DesignWeaver}: {Dimensional} {Scaffolding} for {Text}-to-{Image} {Product} {Design}},
	shorttitle   = {{DesignWeaver}},
	author       = {Tao, Sirui and Liang, Ivan and Peng, Cindy and Wang, Zhiqing and Palani, Srishti and Dow, Steven P.},
	year         = 2025,
	month        = apr,
	booktitle    = {Proceedings of the 2025 {CHI} {Conference} on {Human} {Factors} in {Computing} {Systems}},
	publisher    = {ACM},
	address      = {New York, NY, USA},
	series       = {{CHI} '25},
	pages        = {1--26},
	doi          = {10.1145/3706598.3714211},
	isbn         = {979-8-4007-1394-1},
	url          = {https://dl.acm.org/doi/10.1145/3706598.3714211},
	urldate      = {2025-06-14},
	language     = {en-US}
}

@inproceedings{tholander_design_2023,
	title        = {Design {Ideation} with {AI} - {Sketching}, {Thinking} and {Talking} with {Generative} {Machine} {Learning} {Models}},
	author       = {Tholander, Jakob and Jonsson, Martin},
	year         = 2023,
	month        = jul,
	booktitle    = {Proceedings of the 2023 {ACM} {Designing} {Interactive} {Systems} {Conference}},
	publisher    = {ACM},
	address      = {New York, NY, USA},
	series       = {{DIS} '23},
	pages        = {1930--1940},
	doi          = {10.1145/3563657.3596014},
	isbn         = {978-1-4503-9893-0},
	url          = {https://dl.acm.org/doi/10.1145/3563657.3596014},
	urldate      = {2024-08-29}
}

@inproceedings{uusitalo_clay_2024,
	title        = {”{Clay} to {Play} {With}”: {Generative} {AI} {Tools} in {UX} and {Industrial} {Design} {Practice}},
	shorttitle   = {”{Clay} to {Play} {With}”},
	author       = {Uusitalo, Severi and Salovaara, Antti and Jokela, Tero and Salmimaa, Marja},
	year         = 2024,
	month        = jul,
	booktitle    = {Designing {Interactive} {Systems} {Conference}},
	publisher    = {ACM},
	address      = {IT University of Copenhagen Denmark},
	pages        = {1566--1578},
	doi          = {10.1145/3643834.3661624},
	isbn         = {979-8-4007-0583-0},
	url          = {https://dl.acm.org/doi/10.1145/3643834.3661624},
	urldate      = {2024-07-09},
	language     = {en}
}

@inproceedings{verheijden_collaborative_2023,
	title        = {Collaborative {Diffusion}: {Boosting} {Designerly} {Co}-{Creation} with {Generative} {AI}},
	shorttitle   = {Collaborative {Diffusion}},
	author       = {Verheijden, Mathias Peter and Funk, Mathias},
	year         = 2023,
	month        = apr,
	booktitle    = {Extended {Abstracts} of the 2023 {CHI} {Conference} on {Human} {Factors} in {Computing} {Systems}},
	publisher    = {ACM},
	address      = {Hamburg Germany},
	pages        = {1--8},
	doi          = {10.1145/3544549.3585680},
	isbn         = {978-1-4503-9422-2},
	url          = {https://dl.acm.org/doi/10.1145/3544549.3585680},
	urldate      = {2024-04-24},
	language     = {en}
}

@article{vimpari_adapt-or-type_2023,
	title        = {“{An} {Adapt}-or-{Die} {Type} of {Situation}”: {Perception}, {Adoption}, and {Use} of {Text}-to-{Image}-{Generation} {AI} by {Game} {Industry} {Professionals}},
	shorttitle   = {“{An} {Adapt}-or-{Die} {Type} of {Situation}”},
	author       = {Vimpari, Veera and Kultima, Annakaisa and Hämäläinen, Perttu and Guckelsberger, Christian},
	year         = 2023,
	month        = sep,
	journal      = {Proceedings of the ACM on Human-Computer Interaction},
	volume       = 7,
	number       = {CHI PLAY},
	pages        = {131--164},
	doi          = {10.1145/3611025},
	issn         = {2573-0142},
	url          = {https://dl.acm.org/doi/10.1145/3611025},
	urldate      = {2024-06-18},
	language     = {en}
}

@inproceedings{vogel_conte_2011,
	title        = {Conté: multimodal input inspired by an artist's crayon},
	shorttitle   = {Conté},
	author       = {Vogel, Daniel and Casiez, Géry},
	year         = 2011,
	month        = oct,
	booktitle    = {Proceedings of the 24th annual {ACM} symposium on {User} interface software and technology},
	publisher    = {ACM},
	address      = {Santa Barbara California USA},
	pages        = {357--366},
	doi          = {10.1145/2047196.2047242},
	isbn         = {978-1-4503-0716-1},
	url          = {https://dl.acm.org/doi/10.1145/2047196.2047242},
	urldate      = {2025-09-30},
	language     = {en}
}

@inproceedings{wang_roomdreaming_2024,
	title        = {{RoomDreaming}: {Generative}-{AI} {Approach} to {Facilitating} {Iterative}, {Preliminary} {Interior} {Design} {Exploration}},
	shorttitle   = {{RoomDreaming}},
	author       = {Wang, Shun-Yu and Su, Wei-Chung and Chen, Serena and Tsai, Ching-Yi and Misztal, Marta and Cheng, Katherine M. and Lin, Alwena and Chen, Yu and Chen, Mike Y.},
	year         = 2024,
	month        = may,
	booktitle    = {Proceedings of the {CHI} {Conference} on {Human} {Factors} in {Computing} {Systems}},
	publisher    = {ACM},
	address      = {Honolulu HI USA},
	pages        = {1--20},
	doi          = {10.1145/3613904.3642901},
	isbn         = {979-8-4007-0330-0},
	url          = {https://dl.acm.org/doi/10.1145/3613904.3642901},
	urldate      = {2024-05-13},
	language     = {en}
}

@inproceedings{wang_gentune_2025,
	title        = {{GenTune}: {Toward} {Traceable} {Prompts} to {Improve} {Controllability} of {Image} {Refinement} in {Environment} {Design}},
	shorttitle   = {{GenTune}},
	author       = {Wang, Wen-Fan and Lee, Ting-Ying and Lu, Chien-Ting and Hsu, Che-Wei and Ponsa I Campanyà, Nil and Chen, Yu and Chen, Mike Y. and Chen, Bing-Yu},
	year         = 2025,
	month        = sep,
	booktitle    = {Proceedings of the 38th {Annual} {ACM} {Symposium} on {User} {Interface} {Software} and {Technology}},
	publisher    = {ACM},
	address      = {Busan Republic of Korea},
	pages        = {1--21},
	doi          = {10.1145/3746059.3747774},
	isbn         = {979-8-4007-2037-6},
	url          = {https://dl.acm.org/doi/10.1145/3746059.3747774},
	urldate      = {2025-09-30},
	language     = {en}
}

@inproceedings{wang_aideation_2025,
	title        = {{AIdeation}: {Designing} a {Human}-{AI} {Collaborative} {Ideation} {System} for {Concept} {Designers}},
	shorttitle   = {{AIdeation}},
	author       = {Wang, Wen-Fan and Lu, Chien-Ting and Ponsa i Campanyà, Nil and Chen, Bing-Yu and Chen, Mike Y.},
	year         = 2025,
	month        = apr,
	booktitle    = {Proceedings of the 2025 {CHI} {Conference} on {Human} {Factors} in {Computing} {Systems}},
	publisher    = {ACM},
	address      = {New York, NY, USA},
	series       = {{CHI} '25},
	pages        = {1--28},
	doi          = {10.1145/3706598.3714148},
	isbn         = {979-8-4007-1394-1},
	url          = {https://dl.acm.org/doi/10.1145/3706598.3714148},
	urldate      = {2025-08-18},
	language     = {en-US}
}

@inproceedings{wang_promptcharm_2024,
	title        = {{PromptCharm}: {Text}-to-{Image} {Generation} through {Multi}-modal {Prompting} and {Refinement}},
	shorttitle   = {{PromptCharm}},
	author       = {Wang, Zhijie and Huang, Yuheng and Song, Da and Ma, Lei and Zhang, Tianyi},
	year         = 2024,
	month        = may,
	booktitle    = {Proceedings of the {CHI} {Conference} on {Human} {Factors} in {Computing} {Systems}},
	publisher    = {ACM},
	address      = {Honolulu HI USA},
	pages        = {1--21},
	doi          = {10.1145/3613904.3642803},
	isbn         = {979-8-4007-0330-0},
	url          = {https://dl.acm.org/doi/10.1145/3613904.3642803},
	urldate      = {2024-08-10},
	language     = {en}
}

@inproceedings{weber_drawing--steps_2025,
	title        = {Drawing-in-{Steps}: {Supporting} {Creative} {Goals} through {User} {Engagement} via {Hierarchical} {Image} {Generation}},
	shorttitle   = {Drawing-in-{Steps}},
	author       = {Weber, Christoph Johannes and Huang, Jenny and Rothe, Sylvia},
	year         = 2025,
	month        = may,
	booktitle    = {Proceedings of the 2025 {ACM} {International} {Conference} on {Interactive} {Media} {Experiences}},
	publisher    = {ACM},
	address      = {New York, NY, USA},
	series       = {{IMX} '25},
	pages        = {108--125},
	doi          = {10.1145/3706370.3727862},
	isbn         = {979-8-4007-1391-0},
	url          = {https://dl.acm.org/doi/10.1145/3706370.3727862},
	urldate      = {2025-06-25},
	language     = {en-US}
}

@misc{yu_generative_2018,
	title        = {Generative {Image} {Inpainting} with {Contextual} {Attention}},
	author       = {Yu, Jiahui and Lin, Zhe and Yang, Jimei and Shen, Xiaohui and Lu, Xin and Huang, Thomas S.},
	year         = 2018,
	month        = mar,
	url          = {http://arxiv.org/abs/1801.07892},
	urldate      = {2026-01-15},
	note         = {arXiv:1801.07892}
}

@inproceedings{zamfirescu-pereira_why_2023,
	title        = {Why {Johnny} {Can}’t {Prompt}: {How} {Non}-{AI} {Experts} {Try} (and {Fail}) to {Design} {LLM} {Prompts}},
	shorttitle   = {Why {Johnny} {Can}’t {Prompt}},
	author       = {Zamfirescu-Pereira, J.D. and Wong, Richmond Y. and Hartmann, Bjoern and Yang, Qian},
	year         = 2023,
	month        = apr,
	booktitle    = {Proceedings of the 2023 {CHI} {Conference} on {Human} {Factors} in {Computing} {Systems}},
	publisher    = {ACM},
	address      = {New York, NY, USA},
	series       = {{CHI} '23},
	pages        = {1--21},
	doi          = {10.1145/3544548.3581388},
	isbn         = {978-1-4503-9421-5},
	url          = {https://dl.acm.org/doi/10.1145/3544548.3581388},
	urldate      = {2024-08-29}
}

@inproceedings{zhang_protodreamer_2024,
	title        = {{ProtoDreamer}: {A} {Mixed}-prototype {Tool} {Combining} {Physical} {Model} and {Generative} {AI} to {Support} {Conceptual} {Design}},
	shorttitle   = {{ProtoDreamer}},
	author       = {Zhang, Hongbo and Chen, Pei and Xie, Xuelong and Lin, Chaoyi and Liu, Lianyan and Li, Zhuoshu and You, Weitao and Sun, Lingyun},
	year         = 2024,
	month        = oct,
	booktitle    = {Proceedings of the 37th {Annual} {ACM} {Symposium} on {User} {Interface} {Software} and {Technology}},
	publisher    = {ACM},
	address      = {New York, NY, USA},
	series       = {{UIST} '24},
	pages        = {1--18},
	doi          = {10.1145/3654777.3676399},
	isbn         = {979-8-4007-0628-8},
	url          = {https://dl.acm.org/doi/10.1145/3654777.3676399},
	urldate      = {2024-12-23},
	language     = {en-US}
}

@incollection{cooper_thematic_2012,
	title        = {Thematic analysis.},
	author       = {Braun, Virginia and Clarke, Victoria},
	year         = 2012,
	booktitle    = {{APA} handbook of research methods in psychology, {Vol} 2: {Research} designs: {Quantitative}, qualitative, neuropsychological, and biological.},
	publisher    = {American Psychological Association},
	address      = {Washington},
	pages        = {57--71},
	doi          = {10.1037/13620-004},
	isbn         = {978-1-4338-1005-3},
	url          = {https://content.apa.org/books/13620-004},
	urldate      = {2024-12-20},
	language     = {en},
	editor       = {Cooper, Harris and Camic, Paul M. and Long, Debra L. and Panter, A. T. and Rindskopf, David and Sher, Kenneth J.}
}

@incollection{gero_exploration_2006,
           title = {Exploration through drawings in the conceptual stages of product design},
            year = {2006},
       booktitle = {Design Computing and Cognition},
          editor = {J. Gero},
       publisher = {Springer},
    address      = {Dordrecht},
           month = {July},
           pages = {83--102},
            isbn = {9781402051302},
             url = {https://oro.open.ac.uk/7417/},
          author = {Prats, Miquel and Earl, C. F.}
}

@inproceedings{gaver_ambiguity_2003,
	title        = {Ambiguity as a resource for design},
	author       = {Gaver, William W. and Beaver, Jacob and Benford, Steve},
	year         = 2003,
	month        = apr,
	booktitle    = {Proceedings of the {SIGCHI} {Conference} on {Human} {Factors} in {Computing} {Systems}},
	publisher    = {ACM},
	address      = {New York, NY, USA},
	series       = {{CHI} '03},
	pages        = {233--240},
	doi          = {10.1145/642611.642653},
	isbn         = {978-1-58113-630-2},
	url          = {https://dl.acm.org/doi/10.1145/642611.642653},
	urldate      = {2026-01-20}
}

@inproceedings{landay_interactive_1995,
	title        = {Interactive sketching for the early stages of user interface design},
	author       = {Landay, James A. and Myers, Brad A.},
	year         = 1995,
	month        = may,
	booktitle    = {Proceedings of the {SIGCHI} {Conference} on {Human} {Factors} in {Computing} {Systems}},
	publisher    = {ACM},
	address      = {USA},
	series       = {{CHI} '95},
	pages        = {43--50},
	doi          = {10.1145/223904.223910},
	isbn         = {978-0-201-84705-5},
	url          = {https://dl.acm.org/doi/10.1145/223904.223910},
	urldate      = {2026-01-20},
	note         = {ACM Press/Addison-Wesley Publishing Co.}
}

\appendix

\section{Appendix}

\subsection{System Prompt}
\label{app:system_prompts}
This section presents the system prompt that operationalizes multimodal refinement by mapping user inputs to structured image-editing actions. \autoref{lst:system_prompt} shows the prompt used in the prototype. 

\begin{lstlisting}[
style=python,
caption={System prompt for mapping multimodal user inputs to one of seven predefined refinement actions, appending the corresponding Action Guide, and merging any user-specified ``Required Instructions'' into a single executable editing prompt.},
label={lst:system_prompt}
]
if (mode == "text" or mode == "combined"):
    required_prompt_text =
    """4. Then, if the user has included "Required instructions:", extract the description and append any new information to the selected Action Guide prompt. The result should be a clear and actionable editing instruction that combines the intent of both prompts. For example, if the selected Action Guide prompt is "Add a pattern: Add a wavy pattern to the pillow" and the user's required instructions are "Required instructions: Add a swirly pattern", the final merged prompt should be "Add a pattern: Add a wavy pattern to the pillow. The pattern is swirly."
    """
else:
    required_prompt_text = ""

system_prompt = f"""
You are an expert image-editing prompt engineer and annotation
interpreter.
Your task is to:
1. {mode_prompt} and determine what the user wants to change in the original image.
2. Then classify the user's intended change into one of these seven editing actions:
    1. Add a new object
    2. Change the shape of an object
    3. Add a pattern
    4. Reference the pattern of another object
    5. Change orientation of an object
    6. Change size of an object
    7. Move an object
   Your returned message should begin with one of these seven actions verbatim. For example, if the intended change is moving an object, the output message should start with "Move an object: ...".
3. Then retrieve the corresponding action guide for the identified action from the Action Guide dictionary (provided by the user). Select only one of the seven prompts from the Action Guide.
{required_prompt_text}

Output: Begin with one of the seven editing actions (e.g., "Add a new object:"), then append the selected Action Guide prompt without any changes, followed by additional new information provided by the user or inferred from the input.
Rules:
- Be specific, visual, and instructive (e.g., "Add a large yellow sun with rays in the top-left corner, matching the sketch's placement").
- Do NOT include reasoning or mention of guides or steps---only return the final merged prompt.
- An X mark indicates removing the crossed-out object.
- Arrows combined with the text "pattern" generally indicate referencing the pattern of another object.
- If an object is not mentioned in the required instructions and is not referenced via an arrow or a scribble (e.g., an outline around the object), then it is not part of the edit. In this case, choose a different Action Guide prompt.
"""
return system_prompt
\end{lstlisting}

\subsection{Study Order}
\label{app:study_order}
We counterbalanced the input-modality order using a Latin square with six sequences (\autoref{tab:task_order}).

\begin{table}[!htbp]
    \centering
    \caption{Input-modality order assignment across participants. Thirty participants were assigned to one of six counterbalanced sequences (five participants per sequence).}
    \label{tab:task_order}
    \Description{The table shows the assignment of input-modality orders across participants. Each row corresponds to one of six counterbalanced sequences. Columns indicate the input modality used for Task~1, Task~2, and Task~3 with different image sets, illustrating how text-only, visual-only, and combination modalities were rotated to balance order effects.}
    \small   
    \setlength{\tabcolsep}{4pt}
    \renewcommand{\arraystretch}{1.1}
    {\rowcolors{2}{gray!7}{white}
    \begin{tabularx}{\linewidth}{c >{\centering\arraybackslash}X >{\centering\arraybackslash}X >{\centering\arraybackslash}X}
        \toprule
        \textbf{Sequence} &
        \shortstack[c]{\textbf{Task~1}\\\textit{(Image Set~1)}} &
        \shortstack[c]{\textbf{Task~2}\\\textit{(Image Set~2)}} &
        \shortstack[c]{\textbf{Task~3}\\\textit{(Image Set~3)}} \\
        \midrule
        Sequence~1 & \textonly{Text-only}     & \visualonly{Visual-only}    & \combionly{Combination} \\
        Sequence~2 & \textonly{Text-only}     & \combionly{Combination}     & \visualonly{Visual-only} \\
        Sequence~3 & \combionly{Combination}  & \textonly{Text-only}        & \visualonly{Visual-only} \\
        Sequence~4 & \combionly{Combination}  & \visualonly{Visual-only}    & \textonly{Text-only} \\
        Sequence~5 & \visualonly{Visual-only} & \combionly{Combination}     & \textonly{Text-only} \\
        Sequence~6 & \visualonly{Visual-only} & \textonly{Text-only}        & \combionly{Combination} \\
        \bottomrule
    \end{tabularx}
    }
\end{table}

\subsection{Questionnaire}
\label{app:questionnaire}
This section lists the demographic and background items and describes the study instruments used, including custom scales (Expressing Intent and Iterative Refinement), NASA-TLX subscales, and the UEQ-S.

\subsubsection{Demographics}
\label{app:demographics}
Participants completed a demographic and background questionnaire before the main study. The items below summarize the questionnaire.

\paragraph{Section: Demographic Information}

\aptLtoX{\begin{enumerate}
    \item[1.] What is your gender? \\
    \hspace{1em}$\square$ Male \hspace{1em}$\square$ Female \hspace{1em}$\square$ Non-binary \hspace{1em}$\square$ Prefer not to say \hspace{1em}$\square$ Other
\end{enumerate}}{\begin{enumerate}[label=\arabic*., leftmargin=*]
    \item[1.] What is your gender? \\
    \hspace{1em}$\square$ Male \hspace{1em}$\square$ Female \hspace{1em}$\square$ Non-binary \hspace{1em}$\square$ Prefer not to say \hspace{1em}$\square$ Other
\end{enumerate}}

\paragraph{Section: Design Experience}
\aptLtoX{\begin{enumerate}
    \item[2.] What is your \textbf{current role}? (e.g., UI, UX, interior, graphic, concept designer, design lead)
    \item[3.] What are your \textbf{main tasks} in your current design role?
    \item[4.] Please describe your \textbf{background in design}, including any formal education, training, or work experience.
    \item[5.] How many \textbf{years of experience} do you have in your design field? (Include professional work, internships, or coursework.)
    \item[6.] In which \textbf{country or countries} have you received most of your design education or professional experience? (You may list more than one.)
\end{enumerate}}{\begin{enumerate}[resume, label=\arabic*., leftmargin=*]
    \item[2.] What is your \textbf{current role}? (e.g., UI, UX, interior, graphic, concept designer, design lead)
    \item[3.] What are your \textbf{main tasks} in your current design role?
    \item[4.] Please describe your \textbf{background in design}, including any formal education, training, or work experience.
    \item[5.] How many \textbf{years of experience} do you have in your design field? (Include professional work, internships, or coursework.)
    \item[6.] In which \textbf{country or countries} have you received most of your design education or professional experience? (You may list more than one.)
\end{enumerate}}

\paragraph{Section: Generative AI Experience}
\aptLtoX{\begin{enumerate}
    \item[7.] Have you used generative AI image tools before? \\
    \hspace{1em}$\square$ Yes \hspace{1em}$\square$ No
    \item[7.1] If \textbf{Yes}: Which tools do you currently use or have used often? (e.g., Midjourney, DALL-E, Adobe Firefly, Vizcom)
    \item[7.2] How often do you use generative AI image tools? \\
    \hspace{1em}$\square$ Daily \hspace{1em}$\square$ 2--3 times/week \hspace{1em}$\square$ Weekly \hspace{1em}$\square$ Monthly \hspace{1em}$\square$ Less than once a month
\end{enumerate}}{\begin{enumerate}[resume, label=\arabic*., leftmargin=*]
    \item[7.] Have you used generative AI image tools before? \\
    \hspace{1em}$\square$ Yes \hspace{1em}$\square$ No
    \item[7.1] If \textbf{Yes}: Which tools do you currently use or have used often? (e.g., Midjourney, DALL-E, Adobe Firefly, Vizcom)
    \item[7.2] How often do you use generative AI image tools? \\
    \hspace{1em}$\square$ Daily \hspace{1em}$\square$ 2--3 times/week \hspace{1em}$\square$ Weekly \hspace{1em}$\square$ Monthly \hspace{1em}$\square$ Less than once a month
\end{enumerate}}

\paragraph{Section: Design Tools \& Work Setup}
\aptLtoX{\begin{enumerate}
    \item[8.] What \textbf{software tools} do you use for your design work? (e.g., Figma, Blender, Photoshop, CAD, MS Word)
    \item[9.] What \textbf{input devices} do you typically use for design tasks? (e.g., mouse, touchpad, stylus, Apple Pencil)
    \item[10.] Do you use a \textbf{pen tablet} (i.e., a stylus with a non-display tablet) for your design tasks? \\
    \hspace{1em}$\square$ Yes \hspace{1em}$\square$ No
    \item[10.1] If \textbf{Yes}: How often do you use a pen tablet? \\
    \hspace{1em}$\square$ Daily \hspace{1em}$\square$ 2--3 times/week \hspace{1em}$\square$ Weekly \hspace{1em}$\square$ Monthly \hspace{1em}$\square$ Less than once a month
\end{enumerate}}{\begin{enumerate}[resume, label=\arabic*., leftmargin=*]
    \item[8.] What \textbf{software tools} do you use for your design work? (e.g., Figma, Blender, Photoshop, CAD, MS Word)
    \item[9.] What \textbf{input devices} do you typically use for design tasks? (e.g., mouse, touchpad, stylus, Apple Pencil)
    \item[10.] Do you use a \textbf{pen tablet} (i.e., a stylus with a non-display tablet) for your design tasks? \\
    \hspace{1em}$\square$ Yes \hspace{1em}$\square$ No
    \item[10.1] If \textbf{Yes}: How often do you use a pen tablet? \\
    \hspace{1em}$\square$ Daily \hspace{1em}$\square$ 2--3 times/week \hspace{1em}$\square$ Weekly \hspace{1em}$\square$ Monthly \hspace{1em}$\square$ Less than once a month
\end{enumerate}}

\subsubsection{Expressing Intent Questionnaire (Custom)}
\label{app:expressing-intent_questions}
The following items assessed participants' perceived ability to express refinement intent using the input method.

\begin{enumerate}[leftmargin=*]
    \item \textbf{Overall expression:} The input method allowed me to clearly express all the changes I wanted to make to the image. (Please focus only on the input interaction, not the result images.)
    \item \textbf{Local edits:} The input method allowed me to clearly express small, specific changes to parts of the image (e.g., adjusting a single object's position, shape, size, or style).
    \item \textbf{Visual features and style:} The input method allowed me to clearly express how I wanted objects to look (e.g., their color, shape, pattern, or style).
    \item \textbf{Referencing attributes:} The input method allowed me to clearly refer to features of other objects in the image (e.g., color, pattern, or design).
    \item \textbf{Spatial relationships:} The input method allowed me to clearly express how objects should be placed, moved, or oriented---and how they are positioned relative to each other.
    \item \textbf{Size and scale:} The input method allowed me to clearly express the intended size or scale of objects (including their relative size compared to other objects).
\end{enumerate}

\subsubsection{NASA-TLX Subscales}
\label{app:nasa-tlx}
Five NASA-TLX subscales were used to assess subjective workload across modalities. \textit{Physical Demand} was excluded because physical effort was not a focus of the study.

\begin{enumerate}[leftmargin=*]
    \item \textbf{Mental Demand:} How mentally demanding or confusing was using the input method? (e.g., figuring out how to use it, thinking hard, or navigating menus.)
    \item \textbf{Temporal Demand:} How rushed or time-pressured did you feel while using the input method? (e.g., feeling that you had to act quickly or could not take your time.)
    \item \textbf{Performance:} How successfully were you able to express your design intent using the input method? (Please focus on how clearly you could express your edits, not how the system responded.)
    \item \textbf{Effort:} How much effort did it take to express your design intent using the input method?
    \item \textbf{Frustration:} How insecure, discouraged, irritated, stressed, or annoyed were you with the input method?
\end{enumerate}

\subsubsection{UEQ-S}
\label{app:ueq-s}
The short form of the User Experience Questionnaire (UEQ-S) comprises eight bipolar items that measure pragmatic and hedonic quality using 7-point semantic differential scales (\autoref{tab:ueq-s_items}).

\begin{table}[!htbp]
    \centering
    \caption{UEQ-S items grouped by pragmatic and hedonic quality. Each item is rated on a 7-point semantic differential scale.}
    \label{tab:ueq-s_items}
    \Description{The table lists UEQ-S questionnaire items grouped by quality dimension. Rows are organized into two sections: Pragmatic Quality and Hedonic Quality. Each row presents a bipolar item with opposing adjective anchors shown in the leftmost and rightmost columns. The seven columns between the anchors represent the response options of the semantic differential scale.}
    \scriptsize
    \setlength{\tabcolsep}{4pt}
    \renewcommand{\arraystretch}{1}
    {\rowcolors{2}{gray!7}{white}
    \resizebox{\linewidth}{!}{%
        \begin{tabular}{l*{7}{c}l}
            \toprule
            \multicolumn{9}{c}{\textbf{Pragmatic Quality}} \\[0.2em]
            obstructive      & o & o & o & o & o & o & o & supportive \\
            complicated      & o & o & o & o & o & o & o & easy \\
            inefficient      & o & o & o & o & o & o & o & efficient \\
            confusing        & o & o & o & o & o & o & o & clear \\
            \midrule
            \multicolumn{9}{c}{\textbf{Hedonic Quality}} \\[0.2em]
            boring           & o & o & o & o & o & o & o & exciting \\
            not interesting  & o & o & o & o & o & o & o & interesting \\
            conventional     & o & o & o & o & o & o & o & inventive \\
            usual            & o & o & o & o & o & o & o & leading edge \\
            \bottomrule
        \end{tabular}%
    }}
\end{table}

\subsubsection{Iterative Refinement Questions (Custom)}
\label{app:iterative-refinement_questions}
A custom 7-point Likert-scale questionnaire was used to assess participants' perceptions of the iterative image refinement process across input modalities. The questionnaire consisted of three items capturing perceived ease, satisfaction, and the overall smoothness and effectiveness of multi-step refinement.

\begin{enumerate}[leftmargin=*]
    \item I expect that this input method will make it easy to iteratively and gradually edit my image, step by step.
    \item I expect that this input method will support image refinement over multiple steps in a satisfying way.
    \item I expect that the overall process of iterative editing using this input method will feel smooth and effective.
\end{enumerate}

\subsection{Quantitative Analysis Results}
\label{app:quant_analysis}
This section summarizes inferential statistics for the quantitative measures, including omnibus Friedman tests and Holm-corrected Wilcoxon post-hoc comparisons.

\subsubsection{Friedman Test Results}
\label{app:friedman_results}
Friedman tests revealed overall differences across the three input conditions. For scales with significant omnibus effects, Holm-corrected Wilcoxon signed-rank tests were used for pairwise comparisons. Kendall's $W$ is reported as an effect size for the Friedman tests (see \autoref{tab:friedman_results}).

\begin{table}[!htbp]
    \centering
    \caption{Friedman test statistics for each quantitative scale. All Holm-corrected $p$-values are $< .001$.}
    \label{tab:friedman_results}
    \Description{The table summarizes Friedman test results for four quantitative outcome scales. Rows correspond to scales, and columns report the chi-square statistic, degrees of freedom, Holm-corrected p-value, and Kendall's $W$ effect size.}
    \small
    \setlength{\tabcolsep}{6pt}
    \renewcommand{\arraystretch}{1.05}
    {\rowcolors{2}{gray!7}{white}
    \begin{tabular}{lcccc}
        \toprule
        \textbf{Scale} & $\chi^2$ & $df$ & $p$ & $W$ \\
        \midrule
        NASA-TLX & 18.62 & 2 & \highlight{$< .001$} & .31 \\
        Expressing Intent & 29.04 & 2 & \highlight{$< .001$} & .48 \\
        UEQ-S & 35.84 & 2 & \highlight{$< .001$} & .60 \\
        Iterative Refinement & 27.83 & 2 & \highlight{$< .001$} & .46 \\
        \bottomrule
    \end{tabular}}
\end{table}

\subsubsection{Expressing Refinement Intent}
\label{app:expressing_intent_results}
Pairwise Wilcoxon signed-rank tests revealed widespread statistically significant differences with large effect sizes (\autoref{tab:expressing_intent_all_wilcoxon_comparisons}).

{
\begin{table}[!htbp]
\centering
\caption{Pairwise Wilcoxon signed-rank test results for the \textit{Expressing Intent} scale across modalities. $Z$ is signed by comparison order; $p$ and $|r|$ are unsigned. Holm-corrected significant results are shaded.}
\label{tab:expressing_intent_all_wilcoxon_comparisons}
\Description{The table reports pairwise Wilcoxon signed-rank test results for the Expressing Intent scale across input modalities. Rows are grouped by item (overall expression, local edits, visual features, referencing attributes, spatial relationships, and size and scale). For each item, the table lists three modality comparisons (text-only vs.\ visual-only, text-only vs.\ combination, and visual-only vs.\ combination) and reports the signed $Z$ statistic, uncorrected $p$-value, Holm-corrected $p$-value, and absolute effect size $|r|$. Shaded cells indicate comparisons that remain significant after Holm correction.}
\small
\setlength{\tabcolsep}{3pt}
\renewcommand{\arraystretch}{1.05}

\resizebox{\columnwidth}{!}{%
\begin{tabular}{
    >{\raggedright\arraybackslash}p{2cm}
    l l c c c c
}
\toprule
\textbf{Item} & \textbf{Modality 1} & \textbf{Modality 2} & \textbf{$Z$} & \textbf{$p$} & \textbf{$p_{corrected}$} & \textbf{$|r|$} \\
\midrule

\multirow{3}{2.2cm}{Overall\\Expression}
  & \textonly{Text-only} & \visualonly{Visual-only} & -3.53 & .056 & .056 & .64 \\
  & \textonly{Text-only} & \combionly{Combination} & -3.68 & .002 & \highlight{.005} & .67 \\
  & \visualonly{Visual-only} & \combionly{Combination} & -4.13 & .016 & \highlight{.032} & .76 \\
\addlinespace

\multirow{3}{2.2cm}{Local\\Edits}
  & \textonly{Text-only} & \visualonly{Visual-only} & -3.90 & .006 & \highlight{.012} & .71 \\
  & \textonly{Text-only} & \combionly{Combination} & -4.62 & $< .001$ & \highlight{$< .001$} & .84 \\
  & \visualonly{Visual-only} & \combionly{Combination} & -4.06 & .024 & \highlight{.024} & .74 \\
\addlinespace

\multirow{3}{2.2cm}{Visual\\Features}
  & \textonly{Text-only} & \visualonly{Visual-only} & -3.36 & .006 & \highlight{.013} & .61 \\
  & \textonly{Text-only} & \combionly{Combination} & -4.45 & $< .001$ & \highlight{.001} & .81 \\
  & \visualonly{Visual-only} & \combionly{Combination} & -4.23 & .030 & \highlight{.030} & .77 \\
\addlinespace

\multirow{3}{2.2cm}{Referencing\\Attributes}
  & \textonly{Text-only} & \visualonly{Visual-only} & -4.00 & .011 & \highlight{.032} & .73 \\
  & \textonly{Text-only} & \combionly{Combination} & -3.91 & .011 & \highlight{.032} & .71 \\
  & \visualonly{Visual-only} & \combionly{Combination} & -4.11 & .176 & .176 & .75 \\
\addlinespace

\multirow{3}{2.2cm}{Spatial\\Relationships}
  & \textonly{Text-only} & \visualonly{Visual-only} & -4.35 & $< .001$ & \highlight{$< .001$} & .79 \\
  & \textonly{Text-only} & \combionly{Combination} & -4.58 & $< .001$ & \highlight{$< .001$} & .84 \\
  & \visualonly{Visual-only} & \combionly{Combination} & -3.84 & .398 & .398 & .70 \\
\addlinespace

\multirow{3}{2.2cm}{Size and\\Scale}
  & \textonly{Text-only} & \visualonly{Visual-only} & -3.92 & .003 & \highlight{.007} & .72 \\
  & \textonly{Text-only} & \combionly{Combination} & -4.54 & $< .001$ & \highlight{$< .001$} & .83 \\
  & \visualonly{Visual-only} & \combionly{Combination} & -3.28 & .569 & .569 & .60 \\
\bottomrule
\end{tabular}%
}
\end{table}
}

\subsubsection{Task Completion Time (Efficiency)}
\label{app:task_completion_time}
We compared task completion times across modalities; significant overall differences were observed for Tasks~4--7 (\autoref{tab:task_time_posthoc_wilcoxon}).

{
\begin{table}[!htbp]
\centering
\caption{Post-hoc Wilcoxon signed-rank test results for task completion times. Only Tasks~4--7 showed significant overall effects; shaded cells indicate Holm-corrected significant results.}
\label{tab:task_time_posthoc_wilcoxon}
\Description{The table reports post-hoc Wilcoxon signed-rank test results for task completion times across input modalities. Rows are grouped by task (Tasks~4--7). For each task, three modality comparisons are listed: text-only vs.\ visual-only, text-only vs.\ combination, and visual-only vs.\ combination. Columns report the number of observations, signed $Z$ statistic, uncorrected $p$-value, Holm-corrected $p$-value, and absolute effect size $|r|$. Shaded cells indicate comparisons that remain significant after Holm correction.}
\small
\setlength{\tabcolsep}{3pt}
\renewcommand{\arraystretch}{1.05}

\resizebox{\linewidth}{!}{%
\begin{tabular}{
    >{\raggedright\arraybackslash}p{1.2cm}
    l l c c c c c
}
\toprule
\textbf{Task} & \textbf{Modality 1} & \textbf{Modality 2} & $N$ & $Z$ & $p$ & $p_{corrected}$ & $|r|$ \\
\midrule

\multirow{3}{1.2cm}{Task~4}
  & \textonly{Text-only} & \visualonly{Visual-only} & 30 & -3.82 & $< .001$ & \highlight{$< .001$} & .70 \\
  & \textonly{Text-only} & \combionly{Combination} & 30 & -3.03 & .002 & \highlight{.003} & .55 \\
  & \visualonly{Visual-only} & \combionly{Combination} & 30 & -0.81 & .428 & .428 & .15 \\
\addlinespace

\multirow{3}{1.2cm}{Task~5}
  & \textonly{Text-only} & \visualonly{Visual-only} & 29 & -4.05 & $< .001$ & \highlight{$< .001$} & .75 \\
  & \textonly{Text-only} & \combionly{Combination} & 29 & -3.60 & $< .001$ & \highlight{$< .001$} & .67 \\
  & \visualonly{Visual-only} & \combionly{Combination} & 29 & -1.42 & .162 & .162 & .26 \\
\addlinespace

\multirow{3}{1.2cm}{Task~6}
  & \textonly{Text-only} & \visualonly{Visual-only} & 28 & -3.01 & .002 & \highlight{.006} & .57 \\
  & \textonly{Text-only} & \combionly{Combination} & 28 & -1.80 & .074 & .147 & .34 \\
  & \visualonly{Visual-only} & \combionly{Combination} & 28 & -0.61 & .552 & .552 & .12 \\
\addlinespace

\multirow{3}{1.2cm}{Task~7}
  & \textonly{Text-only} & \visualonly{Visual-only} & 29 & -3.38 & $< .001$ & \highlight{.001} & .63 \\
  & \textonly{Text-only} & \combionly{Combination} & 29 & -1.94 & .053 & .107 & .36 \\
  & \visualonly{Visual-only} & \combionly{Combination} & 29 & -1.55 & .126 & .126 & .29 \\
\bottomrule
\end{tabular}%
}
\end{table}
}

\subsubsection{NASA-TLX}
\label{app:nasa_tlx_tb}
NASA-TLX scores differed significantly across modalities for all items (\autoref{tab:nasa_means_sd}).

\begin{table}[!htbp]
    \centering
    \caption{Mean ($M$) and standard deviation ($SD$) for NASA-TLX items by modality. Lower is better, except for Performance, where higher is better.}
    \label{tab:nasa_means_sd}
    \Description{The table reports descriptive statistics and omnibus test results for NASA-TLX workload items across input modalities. Rows correspond to individual NASA-TLX items. For each modality (text-only, visual-only, and combination), the table lists the mean ($M$) and standard deviation ($SD$). The final columns report Friedman test statistics, including the chi-square value, degrees of freedom, $p$-value, and Kendall's $W$ effect size, summarizing modality differences for each item.}
    \small
    \setlength{\tabcolsep}{4pt}
    \renewcommand{\arraystretch}{1.2}
    {\rowcolors{2}{gray!7}{white}
    \resizebox{\linewidth}{!}{%
        \begin{tabular}{lcccccccccc}
        \toprule
        \textbf{Item} & \multicolumn{2}{c}{\textbf{\textonly{Text-only}}} & \multicolumn{2}{c}{\textbf{\visualonly{Visual-only}}} & \multicolumn{2}{c}{\textbf{\combionly{Combination}}} & \multicolumn{4}{c}{\textbf{Friedman Test}}\\
         \cmidrule(lr){2-3} \cmidrule(lr){4-5} \cmidrule(lr){6-7} \cmidrule(lr){8-11}
         & $M$ & $SD$ & $M$ & $SD$ & $M$ & $SD$ & $\chi^2$ & $df$ & $p$ & $W$\\
        \midrule
        Mental Demand & 3.7 & 1.7 & 2.6 & 1.4 & 2.8 & 1.4 & 7.02 & 2 & \highlight{.030} & .12 \\
        Temporal Demand & 3.1 & 1.5 & 2.5 & 1.5 & 2.4 & 1.3 & 6.19 & 2 & \highlight{.045} & .10 \\
        Performance & 4.1 & 1.3 & 4.8 & 1.4 & 5.6 & 1.0 & 22.92 & 2 & \highlight{$< .001$}  & .38 \\
        Effort & 4.4 & 1.6 & 3.7 & 1.6 & 2.9 & 1.4 & 13.11 & 2 & \highlight{.001} & .22 \\
        Frustration & 3.8 & 1.4 & 2.8 & 1.4 & 2.2 & 1.1 & 22.90 & 2 & \highlight{$< .001$} & .38 \\
        \bottomrule
        \end{tabular}
        }}
\end{table}

\subsubsection{UEQ-S}
\label{app:ueq-s_results}
UEQ-S ratings differed significantly across modalities for pragmatic, hedonic, and overall quality (\autoref{tab:ueqs_all_wilcoxon_comparisons}).

{ 
\begin{table}[!htbp]
\centering
\caption{Post-hoc Wilcoxon signed-rank test results for \textit{UEQ-S} quality dimensions. Holm-corrected significant results are shaded.}
\label{tab:ueqs_all_wilcoxon_comparisons}
\Description{The table reports post-hoc Wilcoxon signed-rank test results for UEQ-S quality dimensions across input modalities. Rows are grouped by quality dimension (pragmatic, hedonic, and overall). For each dimension, three modality comparisons are listed: text-only vs.\ visual-only, text-only vs.\ combination, and visual-only vs.\ combination. Columns report the signed $Z$ statistic, uncorrected $p$-value, Holm-corrected $p$-value, and absolute effect size $|r|$. Shaded cells indicate comparisons that remain significant after Holm correction.}
\small
\setlength{\tabcolsep}{4pt}
\renewcommand{\arraystretch}{1.05}

\resizebox{\linewidth}{!}{%
\begin{tabular}{
    >{\raggedright\arraybackslash}p{1.8cm}
    l l c c c c
}
\toprule
\textbf{Quality Dimension} & \textbf{Modality 1} & \textbf{Modality 2} & \textbf{$Z$} & \textbf{$p$} & \textbf{$p_{corrected}$} & \textbf{$|r|$} \\
\midrule

\multirow{3}{1.8cm}{Pragmatic}
  & \textonly{Text-only} & \visualonly{Visual-only} & -3.57 & .001 & \highlight{.002} & .65 \\
  & \textonly{Text-only} & \combionly{Combination}  & -4.54 & $< .001$ & \highlight{$< .001$} & .83 \\
  & \visualonly{Visual-only} & \combionly{Combination} & -3.51 & .011 & \highlight{.011} & .64 \\
\addlinespace

\multirow{3}{1.8cm}{Hedonic}
  & \textonly{Text-only} & \visualonly{Visual-only} & -4.47 & $< .001$ & \highlight{$< .001$} & .82 \\
  & \textonly{Text-only} & \combionly{Combination} & -4.64 & $< .001$ & \highlight{$< .001$} & .85 \\
  & \visualonly{Visual-only} & \combionly{Combination} & -2.75 & .564 & .564 & .50 \\
\addlinespace

\multirow{3}{1.8cm}{Overall}
  & \textonly{Text-only} & \visualonly{Visual-only} & -4.32 & $< .001$ & \highlight{$< .001$} & .79 \\
  & \textonly{Text-only} & \combionly{Combination} & -4.65 & $< .001$ & \highlight{$< .001$} & .85 \\
  & \visualonly{Visual-only} & \combionly{Combination} & -3.34 & .038 & \highlight{.038} & .61 \\
\bottomrule
\end{tabular}%
}
\end{table}
}

\FloatBarrier
\subsubsection{Iterative Refinement Results}
\label{app:iterative_refinement}
Iterative refinement ratings differed significantly across input modalities (\autoref{tab:iterative_refinement}).

\begin{table}[!htbp]
    \centering
    \caption{Mean ($M$) and standard deviation ($SD$) for iterative refinement ratings by modality. Friedman tests showed significant differences across modalities.}
    \label{tab:iterative_refinement}
    \Description{The table reports descriptive statistics and Friedman test results for iterative refinement ratings across input modalities. Columns report the mean ($M$) and standard deviation ($SD$) for text-only, visual-only, and combination modalities. The final columns report Friedman test statistics, including the chi-square value, degrees of freedom, $p$-value, and Kendall's $W$ effect size.}
    \small
    \setlength{\tabcolsep}{4pt}
    {\rowcolors{2}{gray!7}{white}
    \resizebox{\columnwidth}{!}{%
    \begin{tabular}{lcccccccccc}
    \toprule
    \textbf{Scale} &
    \multicolumn{2}{c}{\textbf{\textonly{Text-only}}} &
    \multicolumn{2}{c}{\textbf{\visualonly{Visual-only}}} &
    \multicolumn{2}{c}{\textbf{\combionly{Combination}}} &
    \multicolumn{4}{c}{\textbf{Friedman Test}} \\
    \cmidrule(lr){2-3} \cmidrule(lr){4-5} \cmidrule(lr){6-7} \cmidrule(lr){8-11}
     & $M$ & $SD$ & $M$ & $SD$ & $M$ & $SD$ & $\chi^2$ & $df$ & $p$ & $W$ \\
    \midrule
    Iterative Refinement & 4.8 & 1.3 & 5.8 & 0.9 & 6.2 & 0.7 & 27.83 & 2 & \highlight{$< .001$} & .46 \\
    \bottomrule
    \end{tabular}
    }}
\end{table}

\clearpage
\onecolumn
\subsection{Distribution of Input Usage within the Combination Modality}
\label{app:strategies_results}
\begin{figure}[h]
  \centering
  \includegraphics[width=.85\linewidth]{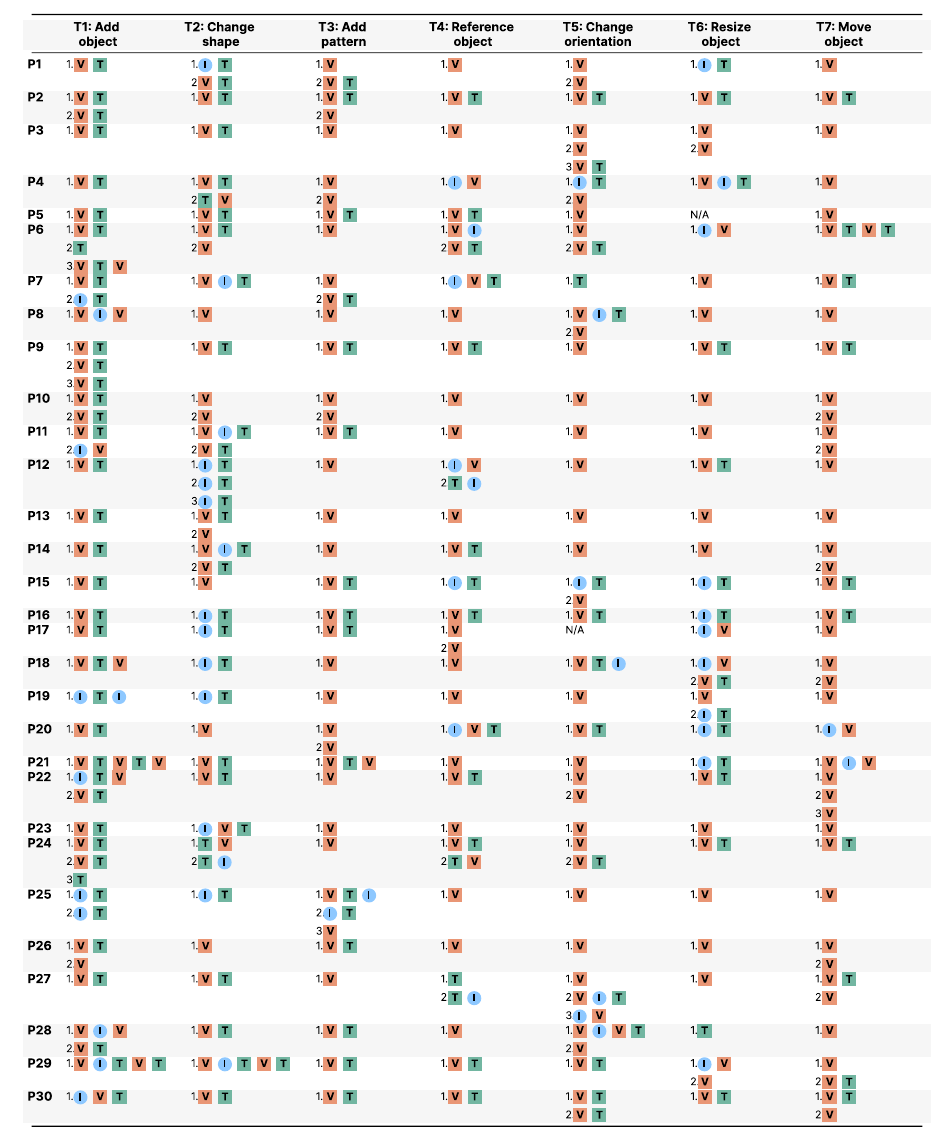}
  \caption{Per-participant modality sequences in the \textbf{Combination} condition, where both text and visual input were available but not required to be used simultaneously. Cells show the left-to-right temporal order of modalities across up to three trials (1--3) per task.  \textbf{T} = text input, \textbf{V} = visual input, \textbf{I} = inpainting; ``N/A'' indicates incomplete trials.}
  \label{fig:modality_use_per_person}
  \Description{The figure shows per-participant input modality sequences in the combination condition across seven refinement tasks. Rows correspond to individual participants, and columns correspond to tasks T1 through T7. Each cell lists the left-to-right order of input modalities used across up to three iterations within a task, including visual input, text input, and inpainting. Abbreviations indicate modality types, and N/A marks incomplete trials. The table provides a detailed overview of how participants combined or alternated input modalities across tasks.}
\end{figure}

This section reports per-participant sequences of input modality use in the \textbf{Combination} condition across seven refinement tasks (\autoref{fig:modality_use_per_person}).

\subsection{User Task Examples}
\label{app:user_input_examples}

\subsubsection{Closed-Ended Tasks}
\begin{figure*}[h]
    \centering
    \includegraphics[width=.92\linewidth]{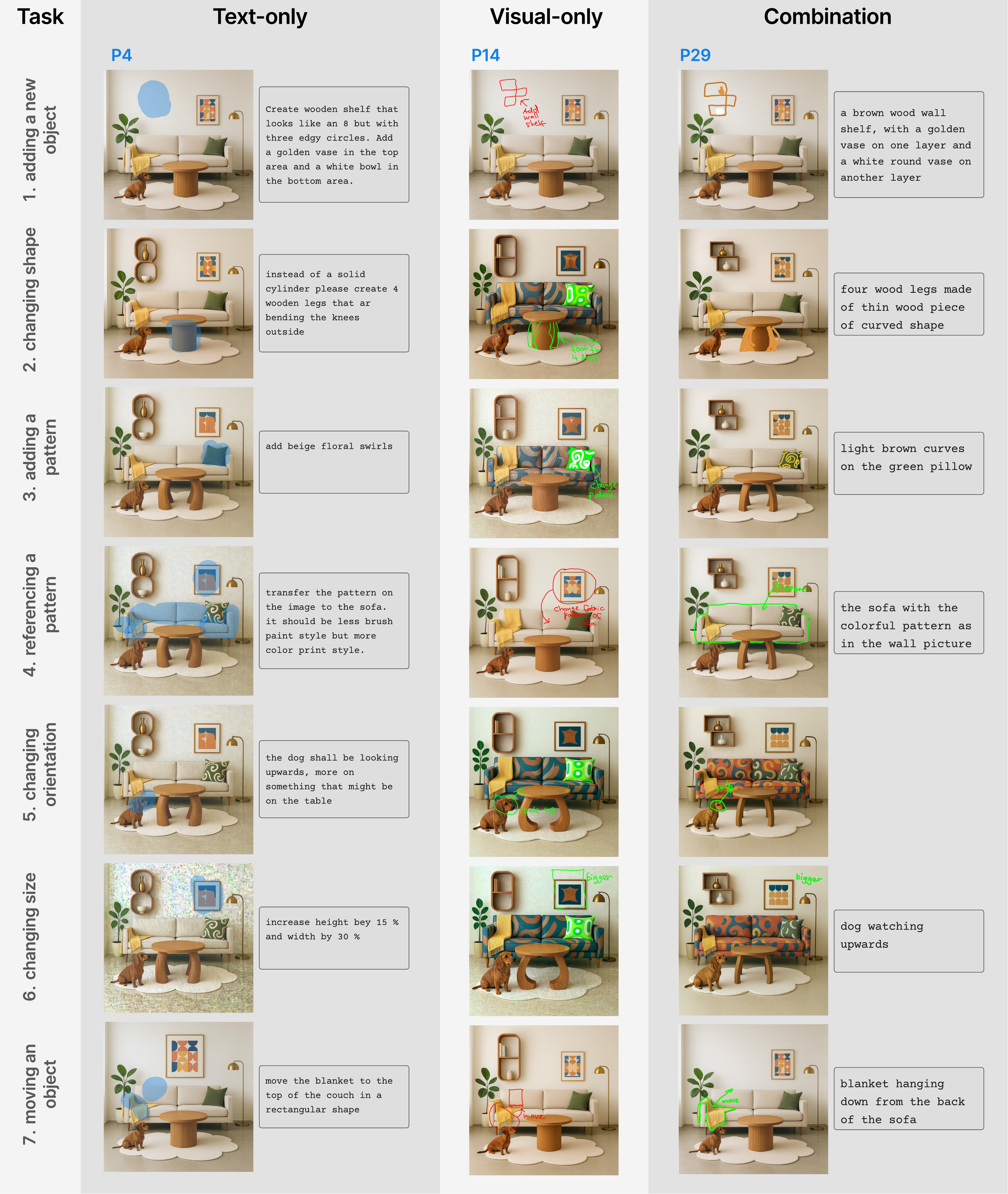}
    \caption{Closed-ended refinement examples from Image Set~1: P4 (\textbf{Text-only}), P14 (\textbf{Visual-only}), and P29 (\textbf{Combination}).}
    \label{fig:examples_closedended}
    \Description{The figure presents closed-ended task examples from a single image set, comparing how the same refinement tasks were expressed using different input modalities. Rows correspond to seven refinement tasks, and columns show examples from three participants using text-only, visual-only, and combination inputs. Each example includes the source image with task-specific annotations or prompts and the resulting refined image. The figure illustrates differences in how identical tasks were specified across input modalities.}
\end{figure*}

This subsection presents representative closed-ended refinement examples, illustrating how the same task was expressed using \textbf{Text-only}, \textbf{Visual-only}, and \textbf{Combination} input (\autoref{fig:examples_closedended}).

\subsubsection{Open-Ended Tasks}
\begin{figure*}[h]
    \centering
    \includegraphics[width=\linewidth]{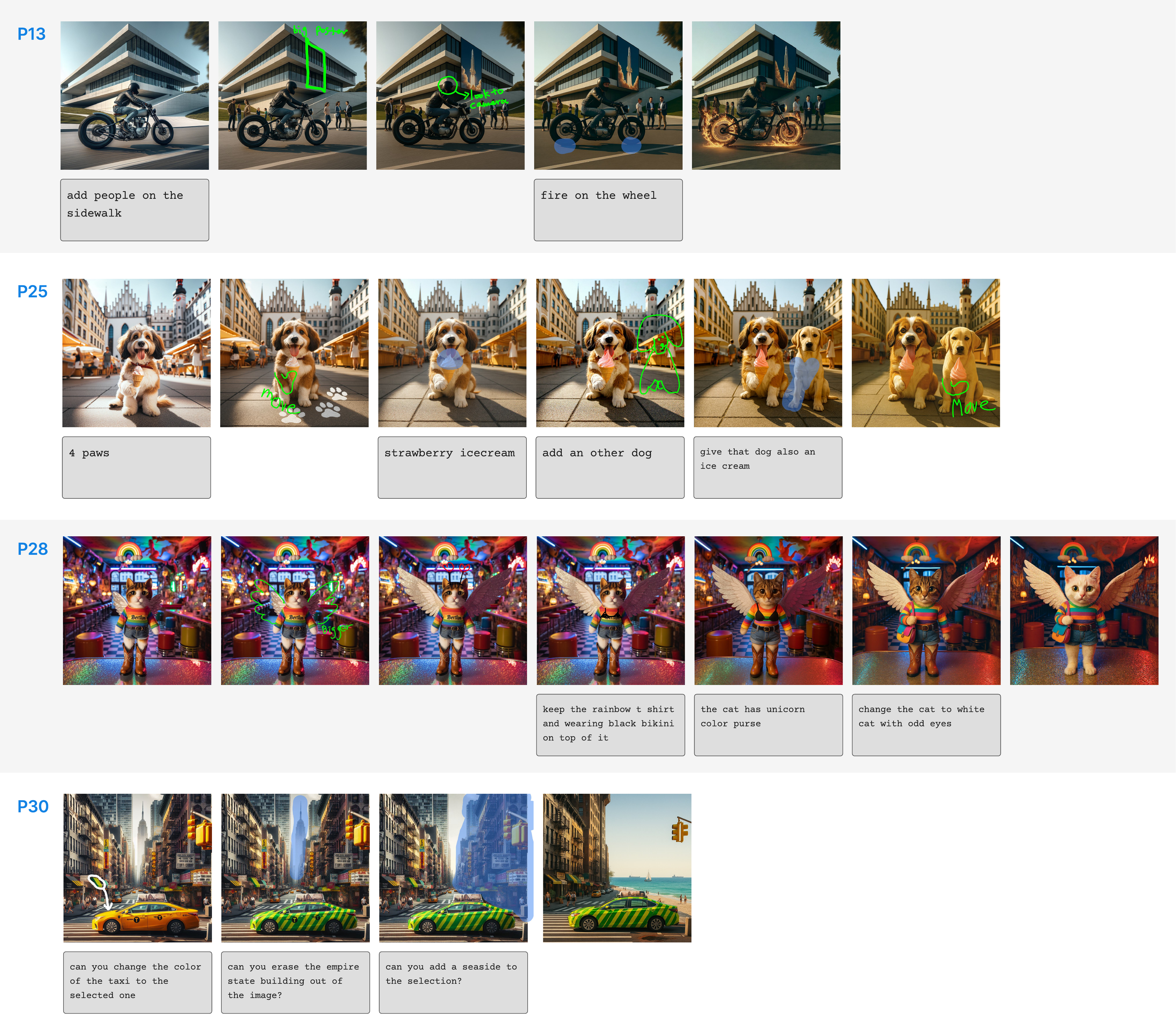}
    \caption{Open-ended refinement examples by P13, P25, P28, and P30, showing participant-driven, iterative refinement without predefined tasks or target images.}
    \label{fig:examples_openended}
    \Description{The figure presents open-ended refinement examples from four participants. Each row shows a sequence of images produced over multiple iterations, starting from an initial generated image and followed by successive refinements. Annotations, scribbles, and text prompts indicate how participants incrementally modified images using their preferred input modalities. The figure illustrates participant-driven, iterative refinement without predefined tasks or target outcomes.}
\end{figure*}

This subsection presents open-ended refinement examples, illustrating how designers generated and refined images without predefined tasks or target images (\autoref{fig:examples_openended}). Participants first generated an initial image and then refined it using their preferred input modalities over multiple iterations.

\twocolumn

\end{document}